\documentclass[sigconf]{acmart}

\def\BibTeX{{\rm B\kern-.05em{\sc i\kern-.025em b}\kern-.08emT\kern-.1667em\lower.7ex\hbox{E}\kern-.125emX}}

\copyrightyear{2019}
\acmYear{2019}
\setcopyright{acmcopyright}
\acmConference[KDD '19]{The 25th ACM SIGKDD Conference on Knowledge
Discovery and Data Mining}{August 4--8, 2019}{Anchorage, AK, USA}
\acmBooktitle{The 25th ACM SIGKDD Conference on Knowledge Discovery and Data
Mining (KDD '19), August 4--8, 2019, Anchorage, AK, USA}
\acmPrice{15.00}
\acmDOI{10.1145/3292500.3330775}
\acmISBN{978-1-4503-6201-6/19/08}
\usepackage{amsmath,amssymb,amsfonts}
\usepackage{algorithm}
\usepackage{algpseudocode}
\usepackage{graphicx}
\usepackage{accents}
\usepackage{multirow}
\usepackage{booktabs}
\usepackage{bm}

\usepackage[tableposition=above]{caption}
\DeclareMathOperator*{\argmin}{\arg\min}

\newcommand{\ouralg}{{BIG-BOSS}} 

\newcommand{\condInd}{{i}}
\newcommand{\flux}{{\bf v}}
\newcommand{\fluxI}{{{\bf v}^{(i)}}}

\newcommand{\fluxMes}{{\bf \tilde{v}}}
\newcommand{\fluxMesI}{{{\bf \tilde{v}}^{(i)}}}

\newcommand{\stoi}{{S}}
\newcommand{\bal}{{\bf b}}
\newcommand{\balI}{{{\bf b}^{(i)}}}
\newcommand{\lowB}{{\bf l}}
\newcommand{\lowBI}{{{\bf l}^{(i)}}}
\newcommand{\uppB}{{\bf u}}
\newcommand{\uppBI}{{{\bf u}^{(i)}}}
\newcommand{\objc}{{\bf c}}
\newcommand{\numMet}{{m}}
\newcommand{\numFlux}{{n}}
\newcommand{\numCond}{{k}}
\newcommand{\numMeas}{{p}}

\newcommand{\newStoi}{{\bf y}}
\newcommand{\newFlux}{{z}}
\newcommand{\newFluxI}{{z^{(i)}}}
\newcommand{\indFI}{{F^{(i)}}}

\newcommand{\dualS}{{\bm \omega}}
\newcommand{\dualSI}{{\bm \omega^{(i)}}}

\newcommand{\dualL}{{\bm \mu}}
\newcommand{\dualLI}{{\bm \mu}^{(i)}}

\newcommand{\dualU}{{\bm \eta}}
\newcommand{\dualUI}{{\bm \eta}^{(i)}}
\newcommand{\objd}{{ d}}
\newcommand{\objdI}{{{ d}^{(i)}}}

\newcommand{\dummy}{{\bf w}}
\newcommand{\dummyI}{{\bf w}^{(i)}}

\newcommand{\sparse}{{\delta}}
\newcommand{\norm}[1]{\left\lVert#1\right\rVert}
\newcommand{\stepsize}{{\alpha}}

\newcommand{\AX}{{\Psi}}
\newcommand{\AZ}{{\Gamma}}
\newcommand{\AU}{{\Xi}}
\newcommand{\AQP}{{\text{QP}}}
\newcommand{\ABI}{{\text{BI}}}
\newcommand{\ABD}{{\text{BD}}}
\newcommand{\ASP}{{\text{SP}}}
\newcommand{\AFX}{{\text{FX}}}
\newcommand{\ANR}{{\text{NR}}}
\newcommand{\ADB}{{\text{DB}}}
\newcommand{\ADE}{{\text{DE}}}
\newcommand{\Arho}{{\rho}}

\newcommand{\Agama}{{\gamma}}
\newcommand{\Aalpha}{{\alpha}}
\newcommand{\AProx}{{\mathbf{Prox}}}

\newcommand{\An}{{{\bf n}}}
\newcommand{\AnI}{{{\bf n}^{(i)}}}

\newcommand{\blkLhsI}{{C^{(i)}}}
\newcommand{\blkRhs}{{{\bf e}}}
\newcommand{\blkRhsI}{{{\bf e}^{(i)}}}
\newcommand{\blkFI}{{P^{(i)}}}
\newcommand{\blkb}{{{\bf q}}}
\newcommand{\blkbI}{{{\bf q}^{(i)}}}
\newcommand{\dualdummy}{{{\bf \lambda}}}
\newcommand{\dualdummyI}{{{\bf \lambda}^{(i)}}}
\newcommand{\scaleUL}{{Q}}
\newcommand{\scaleLR}{{R}}

\newcommand{\T}{^T\!}

\newcommand{\prot}{{\bf p}}
\newcommand{\objp}{{\bf f}}
\newcommand{\pweight}{{\bf a}}
\newcommand{\keff}{{K}}
\newcommand{\slack}{{\bf \sigma}}
\newcommand{\enzyme}{{\bf e}}
\newcommand{\cplxstoi}{{E}}
\newcommand{\ptot}{{t}}

\newcommand{\objpc}{\bar{\objc}}
\newcommand{\pcbal}{\bar{\bal}}
\newcommand{\pclowB}{\bar{\lowB}}
\newcommand{\pcuppB}{\bar{\uppB}}
\newcommand{\pcstoi}{\bar{\stoi}}
\begin{document}

%
% The "title" command has an optional parameter, allowing the author to define a "short title" to be used in page headers.
\title[Estimating Cellular Goals from High-Dimensional Biological Data]{Estimating Cellular Goals from\\ High-Dimensional Biological Data}

% \icmlaffiliation{ucsd}{Department of Bioengineering, 
% University of California at San Diego, La Jolla, CA, USA}
% \icmlaffiliation{bc}{Computer Science Department,
% Boston College, Chestnut Hill, MA, USA}
% \icmlaffiliation{qc}{D\'{e}partment de Biologie,
% Universit\'{e} de Sherbrooke, Sherbrooke, Qu\'{e}bec, Canada}
% \icmlaffiliation{stan}{Department of Management Science and Engineering,
% Stanford University, Stanford, CA, USA}
% \icmlaffiliation{cfb}{The Novo Nordisk Foundation Center 
% for Biosustainability, Technical University of Denmark, Lyngby, Denmark}

% The "author" command and its associated commands are used to define the authors and their affiliations.
% Of note is the shared affiliation of the first two authors, and the "authornote" and "authornotemark" commands
% used to denote shared contribution to the research.
\author{Laurence Yang}
%\author{Bernhard O. Palsson}
\authornote{Both authors contributed equally to this project and are corresponding authors.}
\email{lyang@eng.ucsd.edu}
%\email{palsson@ucsd.edu}
\affiliation{%
\institution{University of California, San Diego}
  %\streetaddress{9500 Gilman Drive}
  %\city{La Jolla}
  %\state{CA}
  %\postcode{92093-0412}
  %\country{USA}
}

\author{Michael A. Saunders}
\email{saunders@stanford.edu}
\affiliation{%
  \institution{Stanford University}
%   \streetaddress{475 Via Ortega}
%   \city{Stanford}
%   \state{CA}
%   \postcode{94305-4042}
}

\author{Jean-Christophe Lachance}
\email{jelachance@eng.ucsd.edu}
\affiliation{%
  \institution{Universit\'{e} de Sherbrooke}
%   \city{Sherbrooke}
%   \state{Qu\'{e}bec}
%   \country{Canada}
}

\author{Bernhard O. Palsson}
\email{palsson@ucsd.edu}
\affiliation{%
\institution{University of California, San Diego}
%   \streetaddress{9500 Gilman Drive}
%   \city{La Jolla}
%   \state{CA}
%   \postcode{92093-0412}
}

\author{Jos\'{e} Bento}
\authornotemark[1]
\email{jose.bento@bc.edu}
\affiliation{%
  \institution{Boston College}
%  \streetaddress{P.O. Box 1212}
%   \city{Chestnut Hill}
%   \state{MA}
%  \country{USA}
}

% \author{Bernhard O. Palsson}
% \affiliation{%
%  \institution{University of California, San Diego}
%  \streetaddress{9500 Gilman Drive}
%  \city{La Jolla}
%  \state{CA}
%  \postcode{92093-0412}
%  \country{USA}
%  }

%
% By default, the full list of authors will be used in the page headers. Often, this list is too long, and will overlap
% other information printed in the page headers. This command allows the author to define a more concise list
% of authors' names for this purpose.
\renewcommand{\shortauthors}{Yang and Bento, et al.}

% \icmlauthor{Laurence Yang}{ucsd}
% \icmlauthor{Jos\'{e} Bento}{bc}
% \icmlauthor{Jean-Christophe LaChance}{qc}
% \icmlauthor{Michael A. Saunders}{stan}
% \icmlauthor{Bernhard O. Palsson}{ucsd,cfb}
% \end{icmlauthorlist}

% \icmlaffiliation{ucsd}{Department of Bioengineering, 
% University of California at San Diego, La Jolla, CA, USA}
% \icmlaffiliation{bc}{Computer Science Department,
% Boston College, Chestnut Hill, MA, USA}
% \icmlaffiliation{qc}{D\'{e}partment de Biologie,
% Universit\'{e} de Sherbrooke, Sherbrooke, Qu\'{e}bec, Canada}
% \icmlaffiliation{stan}{Department of Management Science and Engineering,
% Stanford University, Stanford, CA, USA}
% \icmlaffiliation{cfb}{The Novo Nordisk Foundation Center 
% for Biosustainability, Technical University of Denmark, Lyngby, Denmark}

%%%%% Cellular  GOAL not OBJECTIVE (confusing with optimization  objective)
\begin{abstract}
Optimization-based models have been used to predict cellular behavior
%the operation of cell metabolism 
for over 25 years.
The constraints in these models are derived from genome annotations, measured macro-molecular composition of cells, 
and by measuring
the cell's growth rate and metabolism in different conditions.
The \emph{cellular goal} (the optimization problem that the cell is trying to solve) can be challenging to derive experimentally for many organisms, including human or mammalian cells, which have complex metabolic capabilities and are not well understood.
Existing approaches to learning goals from data include (a) estimating a linear objective function, or (b) estimating linear constraints that model complex biochemical reactions and constrain the cell's operation.
The latter approach is important because often the known/observed biochemical reactions are not enough to explain observations, and hence there is a need to extend automatically the model complexity by learning new chemical reactions. However, this leads to nonconvex optimization problems, and existing tools cannot scale to realistically large metabolic models.
Hence, constraint estimation is still used sparingly despite its benefits for modeling cell metabolism, which is important for developing novel antimicrobials against pathogens, discovering cancer drug targets, and producing value-added chemicals.
Here, we develop the first
approach to estimating constraint reactions from data that can scale to realistically large metabolic models. Previous tools have been used on problems having less than 75 biochemical reactions and 60 metabolites, which limits real-life-size applications.
We perform extensive experiments using 75 large-scale metabolic network models for different organisms (including bacteria, yeasts, and mammals) and show that our algorithm can recover cellular constraint reactions.  The recovered constraints enable accurate prediction of metabolic states in hundreds of growth environments not seen in training data, and we recover useful cellular
goals even when some measurements are missing.
\end{abstract}

% <concept_id>10003752.10003809.10003716.10011138.10011140</concept_id>
% <concept_desc>Theory of computation~Nonconvex optimization</concept_desc>
% <concept_id>10003752.10003809.10010172</concept_id>
% <concept_desc>Theory of computation~Distributed algorithms</concept_desc>
% <concept_id>10010405.10010444.10010087</concept_id>
% <concept_desc>Applied computing~Computational biology</concept_desc>
% <concept_id>10010405.10010444.10010095</concept_id>
% <concept_desc>Applied computing~Systems biology</concept_desc>
% \ccsdesc[500]{Theory of computation~Nonconvex optimization}
% \ccsdesc[500]{Theory of computation~Distributed algorithms}
% \ccsdesc[500]{Applied computing~Computational biology}
% \ccsdesc[500]{Applied computing~Systems biology}

%
% Keywords. The author(s) should pick words that accurately describe the work being
% presented. Separate the keywords with commas.

% %
% % A "teaser" image appears between the author and affiliation information and the body 
% % of the document, and typically spans the page. 
%   \Description{Enjoying the baseball game from the third-base seats. Ichiro Suzuki preparing to bat.}

%
% This command processes the author and affiliation and title information and builds
% the first part of the formatted document.
\maketitle

%----------------------------------------------------------
\section{Introduction and Related Work}
%----------------------------------------------------------

%Systems biology is a disciple that aims to achieve a 
%system-level understanding of organisms, 
%thereby facilitating advancement of biotechnology and 
%medicine \cite{ghosh2011software}.
%This task often entails discovering meaningful
%relationships between process variables of 
%complex biological networks, typically from large-scale
%measurements originating from disparate biological processes
%operating at multiple length and time scales.
%Such measurements are collectively referred to as
%``omics'' as they involve measuring the complete
%make up of a given biological variable (e.g., proteomics 
%attempts to measure the complete protein composition
%of a cell).

%**************************************************
% Need engineers to be inspired by the practical
% importance of this problem and its utility if solved
%**************************************************
Engineered microbial and mammalian cells are used 
as production platforms to synthesize commodity or 
specialty chemicals and pharmaceuticals.
%,often from sustainable feedstocks. 
%
Accurate computational models of cell metabolism and protein
expression are important to
design cell factories and to maximize product yield and
%quality \cite{yimBDO,hefzi2016consensus,kuo2018emerging}. 
quality \cite{yimBDO,kuo2018emerging}. 
Similarly, mathematical models of cell metabolism have been
used to identify strategies to improve the efficacy of
existing antibiotics \cite{brynildsen2013ros}.

The ability of engineers to predict microbial behavior
was facilitated in 1990 by the observation that overflow
metabolism---an industrially-relevant metabolic behavior%
%the production of seemingly wasteful acetate despite the availability of efficient aerobic respiration for energy production
---in \textit{Escherichia coli} could be
predicted by a relatively simple network of reactions with capacity-constrained
flows
\cite{majewski90}.
Since then, this constrained optimization model of
cellular metabolism (often called COBRA) has been applied to
over $78$ species across the tree of life \cite{monkNBT2014}.
Metabolic reconstructions today are 
``genome-scale''---i.e., they account for the majority of 
metabolic genes annotated in the organism's genome---and 
consist of hundreds to thousands
of biochemical reactions and metabolites. For example, 
the most recent reconstruction of 
human metabolism includes 13,543 metabolic reactions
involving 4,140 metabolites \cite{brunk2018recon3d}, 
%\textit{E. coli} metabolism \cite{iML1515} includes 
%2,719 reactions involving 1,192 metabolites, 
while the latest multiscale model of metabolism and 
protein expression for \textit{E. coli} 
\cite{lloyd2018cobrame} consists of 
12,655 reactions and 7,031 components including 
macromolecules like protein, RNA (ribonucleic acid), and ribosome.
%human metabolism  includes 
%reactions and {Y} metabolites.

As of mid-2013, over 640 published studies used COBRA 
for experimental investigation in various domains of
engineering and health sciences \cite{bordbarCOBRA14}.
Successful applications of COBRA include
engineering microbes to produce commodity or valuable
chemicals \cite{yimBDO}, developing novel 
antimicrobials against infectious disease 
%\cite{kim2011integrative,brynildsen2013ros},
\cite{brynildsen2013ros},
and discovering new drug targets against cancer 
%\cite{jerby2010computational,frezza2011haem,folger2011predicting}.
%\cite{frezza2011haem,folger2011predicting}.
\cite{folger2011predicting}.

% SHOW BASIC FORMULATION
In its basic form, COBRA predicts $\numFlux$ reaction fluxes $\flux \in \mathbb{R}^\numFlux$ (reaction rate normalized by
dry weight of biomass) in a 
metabolic network, consisting of $\numMet$ metabolites and $\numFlux$ reactions,
by solving a linear program that models cellular goal:
\begin{equation}\label{eq:FBA}
\max_{\flux} \objc\T \flux \text{\ subject to\ } \stoi \flux = \bal ,\  \lowB \leq \flux \leq \uppB,
\end{equation}
where $\objc\in \mathbb{R}^\numFlux$ is a vector of objective coefficients,
$(\cdot)^T$ denotes the transpose,
$\stoi \in \mathbb{R}^{\numMet \times \numFlux}$ is a matrix of stoichiometric coefficients (one column per biochemical reaction),
$\bal \in \mathbb{R}^\numMet$ is a vector of metabolite accumulation 
or depletion rates,
and $\lowB , \uppB \in \mathbb{R}^\numFlux$ are lower and upper flux bounds.

In  order to make accurate predictions, these models 
require an accurate $\stoi$ matrix, which is nowadays 
reliably reconstructed from extensive knowledgebases of
enzyme biochemistry, and genome annotations for thousands of 
species and strains organisms.
The other key ingredient is the cellular objective function,
$\objc\T \flux$.
Other objectives, including nonlinear functions,
have been tested \cite{schuetz07}.
For many microbes cultured in nutrient-limiting conditions, 
maximization of cell growth rate
 (per unit of limiting nutrient) 
is an accurate objective 
function \cite{schuetz07}.
This particular function corresponds to a $\objc$ that is an indicator vector with a $1$ in the component associated with the reaction
that corresponds to the cell growth, and zero everywhere else.

%Currently, identifying useful objective functions to predict
%the behavior of non-growing microbes or eukaryotes and
%mammalian cells remains an important challenge.

Currently, system-level understanding of living organisms
requires the analysis of large-scale
measurements originating from disparate biological processes
operating at multiple length and time scales.
Such measurements are collectively referred to as
``omics'', as they involve measuring the complete
makeup of a given biological variable (e.g., proteomics 
attempts to measure the complete protein composition
of a cell). Analysis of such omics measurements has shown that,
despite the complexity inherent in living systems, relatively simple models are accurate enough for several biological studies.
%approximations may be useful for modeling them.
E.g., the gene expression profiles of various cell types
can be described in terms of relatively few
biological functions \cite{hartTasks2015}.

Similarly, the metabolic fluxes in a cell can be 
predicted by assuming that the cell is solving an optimization problem shaped by its evolutionary history \cite{feist2016cells}.
The problem includes constraints that model biochemical ``reactions''
consuming and producing metabolites at specific stoichiometric ratios, and an objective that depends on the fluxes through the reactions.
For example, the metabolism of fast-growing microbes (mid-log phase) is predicted accurately by a linear problem such as \eqref{eq:FBA}.
As another example, the metabolism of mammalian cells (e.g., hybridoma), 
is explained well by minimization of the flux through the reaction that makes reactive oxygen species \cite{feist2016cells}.
Studies have identified alternative objective functions
that best predict microbial metabolism under
different growth conditions \cite{schuetz07}. These objectives
include maximizing ATP yield per flux unit and maximizing ATP or biomass yield.

%--------------------------------------------
% ******************************************
While the aforementioned studies have focused on using pre-defined optimization problems, a number of studies have investigated data-driven estimation of cellular goals.
Two types of method exist.
Methods of the first type estimate how important the different chemical reactions are for the cell's operation, i.e.,  estimate $\objc$ in \eqref{eq:FBA}. For example, ObjFind \cite{objfind} does this through a nonconvex optimization
formulation, and the more recent invFBA
%\cite{invFBA,zhao2015learning}
\cite{invFBA}
solves a linear program.
The second type, and the focus of our work, estimate the 
stoichiometric coefficients of a new
chemical reaction, i.e., estimate a new column for matrix $\stoi$ in \eqref{eq:FBA}. 
This approach is important because often the known biochemical reactions are not enough to explain observed data. We want to extend the model complexity automatically by learning new chemical reactions, i.e., new columns for $\stoi$.

%--------------------------------------------

%since it can identify \textit{de novo} objectives 
%that may not be describable by a linear combination of 
%existing reactions in the metabolic reconstruction.
%
%
The main drawback of estimating new reactions is that it requires solving nonconvex optimization problems. Currently, only the BOSS tool
\cite{boss2008} does this. 
BOSS was shown to recover known biomass reactions successfully in synthetic experiments involving less than $70$ reactions and $60$ metabolites.  This is not large enough for real-life applications, which can involve thousands of reactions/metabolites.
Note that BOSS (a) uses an off-the-shelf solver (e.g. MINOS) that cannot exploit parallelism, (b) cannot induce sparsity on the learnt reaction, and (c) has not been tested on large problems.

Here we address these important limitations. Our main contributions are the following.
% metabolic objective search 
\begin{enumerate}
    \item We develop BIG-BOSS (\url{https://github.com/laurenceyang33/cellgoal}), the first tool that can learn biochemical reactions in large models (with up to 4,456 metabolites and 6,663 reactions). %iCHOv1 model
    \item BIG-BOSS is based on the alternating direction method of multipliers (ADMM). It employs an adaptive penalty and a preconditioning scheme, and hence requires no tuning. Further, it can exploit multiple CPUs to speed up computations on large problems. 
    \item BIG-BOSS uses a sparsity-inducing regularization
    to allow users to control the complexity of inferred
    %biochemical
    reactions.
    \item We test and report the accuracy of BIG-BOSS on 75 genome-scale models of cell metabolism including prokaryotes, eukaryotes, and mammalian cells, using both clean and corrupted input datasets.
\end{enumerate}
%

%In this work, we build upon BOSS \cite{boss2008}, and extend the approach of \textit{de novo} objective  reaction generation to include sparsity of the objective stoichiometry, and inferring an objective from multiple conditions. We demonstrate 
% the latest genome-scale model of 
%\textit{E. coli} \cite{iML1515}.

\section{Reaction estimation as an optimization problem}
We estimate a cellular goal using measured fluxes (cf.\ $\flux$ in \eqref{eq:FBA}) as input. These fluxes are measured for different cell growth environments, e.g., 
growth on glucose, or with/without oxygen.
The part of the cellular goal that we estimate here is one new metabolic reaction,
in particular 
its stoichiometric coefficients. In other words, we want to learn a new column for $\stoi$ in \eqref{eq:FBA} in order to explain observations better.
We focus on one reaction for simplicity, but BIG-BOSS can be used to estimate simultaneously the  stoichiometric coefficients of multiple new biochemical reactions.

\subsection{Bilevel optimization problem}

We formulate our problem as a
bilevel optimization problem \cite{bard2013practical}. We start from fluxes
$\{\fluxMesI\}^\numCond_{i=1}$ measured across $\numCond$ growth conditions,
% [LY] 05/16/2019:
which is accomplished by feeding an organism carbon-13 labeled nutrients 
and tracing their distribution using spectroscopy/spectrometry 
\cite{antoniewicz2015methods}.
Each vector $\fluxMesI$ 
contains the fluxes through a subset of all $\numFlux$ 
reactions taking place in the cell. We also fix the flux $\newFluxI \geq 0, \newFluxI \in \mathbb{R}$, through the new undefined reaction whose stoichiometric coefficients 
$\newStoi \in \mathbb{R}^\numMet$ we want to estimate. Given a set of 
coefficients $\newStoi$, 
and for each growth condition $\condInd$, we assume that the cell fluxes 
$\fluxI \in \mathbb{R}^\numFlux$ optimize a standard COBRA model:
\begin{align}
   \fluxI \in \mathcal{C}^{(i)}(\newStoi) \triangleq \arg\max_{\dummyI}
                     \quad & \objc\T \dummyI + \objdI \newFluxI          \label{eq:inner_FBA}
\\ \text{subject to} \quad & \stoi \dummyI + \newFluxI \newStoi = \balI, \label{eq:inner_FBA_constraint_1}
\\                         &  \lowBI \leq \dummyI \leq \uppBI,           \label{eq:inner_FBA_constraint_2}
\end{align}
where $\lowBI$ and $\uppBI$ (known) are the lower and upper flux bounds, $\balI$ (known) is the rate of accumulation or depletion of different metabolites, $\objc$ (known) is the relative importance of the different known chemical reactions in the model, 
$\objdI \in \mathbb{R}$ (known) is the relative importance of the new chemical reaction we want to learn,
and $S$ (known) are the stoichiometric coefficients of the known chemical reactions in the model.

Our goal is to find sparse reaction coefficients $\newStoi$ (unknown) for the new reaction, such that the resulting 
fluxes $\{\fluxI\}^\numCond_{i=1}$
explain the measured fluxes 
$\{\fluxMesI \}^\numCond_{i=1}$ 
as well as possible. For each growth condition $\condInd$,
we keep track of which of the $\numFlux$ fluxes we measure 
in a binary matrix $\indFI$ that, when applied to 
$\fluxI$, selects the correct measured components. 
This leads to the formulation that BIG-BOSS solves:
\begin{align}
\min_{\{\fluxI\}^\numCond_{i=1},\newStoi} \ & \frac{1}{\numCond}
    \sum_{i=1}^\numCond \norm{ \indFI \fluxI - \fluxMesI }_2^2 + 
    \sparse \norm{ \newStoi }_1 \label{eq:biLevelProg}\\
\text{ subject to} \ & \fluxI \in \mathcal{C}_i(\newStoi), \nonumber
\end{align}
where $\sparse \geq 0$ controls sparsity. Larger $\sparse$ encourages reactions having few reactants and products.

We reformulate this bilevel optimization problem as a single-level optimization problem.
\begin{theorem}\label{th:bilevel_to_singlelevel}
Problem \eqref{eq:biLevelProg}, can be written as 
\begin{align}
   \min_{\{\fluxI\},\newStoi,\{\dualSI\},\{\dualLI\},\{\dualUI\}} \quad
   &\frac{1}{\numCond}\sum^\numCond_{i=1}
    \norm{\indFI \fluxI-\fluxMesI}^2_2 + \sparse \norm{\newStoi}_1 \label{eq:single_level_inverse_FBA}
\\ \text{subject to} \quad \qquad \qquad \qquad
   &\hspace{-55pt} S\fluxI + \newFluxI \newStoi  = \bal, \forall i,\label{eq:bilinear_Li_1}
\\ &\hspace{-55pt} \stoi\T \dualSI - \dualLI + \dualUI = \objc, \forall i,\label{eq:bilinear_Li_2}
\\ &\hspace{-55pt} \newStoi\T \dualSI  \geq  \objdI ,\forall i, \label{eq:bilinear_constraint}
\\ &\hspace{-55pt} \objc\T \fluxI + \newFluxI = 
      \bal\T \dualSI - \lowBI\T \dualLI + \uppBI\T \dualUI,\forall i, \label{eq:dual_equals_primal}
\\ &\hspace{-55pt} \lowBI \leq \fluxI \leq \uppBI,\forall i, \label{eq:bilinear_Li_3}
\\ &\hspace{-55pt} \dualLI, \dualUI \geq 0,\forall i,\label{eq:bilinear_Li_4}
\end{align}
where $\dualSI$ is the dual variable for constraint \eqref{eq:inner_FBA_constraint_1}, and $\dualLI$ and $\dualUI$ are dual variables for the upper and lower bounds in \eqref{eq:inner_FBA_constraint_2}.
\end{theorem}

The proof of Theorem \ref{th:bilevel_to_singlelevel} follows the same ideas as in \cite{boss2008}. In a nutshell, 
we write the Karush-Kuhn-Tucker (KKT) conditions for the convex problem \eqref{eq:inner_FBA}, 
and add constraint \eqref{eq:dual_equals_primal} to say that the primal and dual objective values for \eqref{eq:inner_FBA} are equal, which is true by strong duality.
%We include a self-contained proof of Theorem \ref{th:bilevel_to_singlelevel} in Appendix \ref{app:bilevel_proof}.

%The resulting single-level optimization problem is
%
% \begin{align}
% \min_{v,y,w,\mu,\eta}  \quad & \|v-\tilde{v} \|^2_2 + \delta \| y \|_1 \\
% \text{subject to} \quad & Sv + zy = b \\
%                   & S^T \omega - \mu + \eta = c \\
%                   & y^T \omega \geq d \\
%                   & c^T v + z = b^T \omega - l^T \mu + u^T \eta \\
%                   & l \leq v \leq u \\
%                   & \mu, \eta \geq 0,
% \end{align}

The bilinear constraints \eqref{eq:bilinear_constraint} make
the problem nonconvex.
Furthermore, all of the constraints are coupled by the variable $\newStoi$, the new chemical reaction that we want to learn. If it were not for \eqref{eq:bilinear_constraint}, 
we could decouple problem \eqref{eq:single_level_inverse_FBA} into independent generalized LASSO problems \cite{xu2017admm}.
To solve \eqref{eq:single_level_inverse_FBA}--\eqref{eq:bilinear_Li_4}, we use the Alternating Direction Method of Multipliers (ADMM)
\cite{derbinsky2013improved, bento2013message, boyd2011distributed}.
%
%\subsection{Nomenclature}
%Capital letters for matrices.
%Bold letters for vectors.
%Normal letters for primal variables.
%Greek letters for dual variables.
%Known vs est
%input output
%importance/relevance

% Below only needed for arxiv
%\vspace*{-1.5cm}
\section{Summary of the notation used}
All of the variables and parameters that we use are defined in Table~\ref{tab:nomenclature}.
Vectors are written in bold, matrices are capitalized, and
scalars are lower-case.

A few variables are defined locally, and their definitions only hold inside the section, or subsection, where they are defined. These are the only variables not listed in the notation tables. 

%%% Separate line for model variables
%%% ADMM Variables
\begin{table}[t]   % Table 1
\centering
\caption{Nomenclature of variables and parameters \label{tab:nomenclature}}
\begin{tabular}{ccp{1.93in}}
    \toprule
    Variable & Space & Description \\
    \midrule
    \multicolumn{3}{c}{\emph{Model variables}}\\
    \midrule
    \numFlux & $\mathbb{N}$  & number of fluxes in the cell\\
    \numMet & $\mathbb{N}$  & number of metabolites in the cell \\
    \numCond & $\mathbb{N}$  & number of experimental conditions \\
    \numMeas & $\{1,\dots,\numFlux\}$  & number of measured fluxes \\
    $i$ & $\{1,\dots,k\}$  & condition $i$ \\
    $\flux$, $\fluxI$ & $\mathbb{R}^\numFlux$  & all fluxes in model \\
    $\fluxMes$, $\fluxMesI$ & $\mathbb{R}^\numMeas$ & measured fluxes \\
    $\objc$ & $\mathbb{R}^\numFlux$ & relative importance of reaction \\
    $\stoi$ & $\mathbb{R}^{\numMet\times \numFlux}$ & stoichiometric coefficients \\
    $\bal$, $\balI$ & $\mathbb{R}^\numMet$ & rate of accumulation or depletion of metabolites                             (typically zero) \\
    $\lowB$, $\lowBI$ & $\mathbb{R}^\numFlux$ & flux lower bounds \\
    $\uppB$, $\uppBI$ & $\mathbb{R}^\numFlux$ & flux upper bounds \\
    $\newStoi$ & $\mathbb{R}^\numMet$ & coefficients of the new reaction, which we want to learn from data\\
    $\newFlux$, $\newFluxI$ & $\mathbb{R}$   & fixed flux through new reaction \\
    $\dualS$, $\dualSI$ & $\mathbb{R}^\numMet$ & dual variables for linear constraints \\
    $\dualL$, $\dualLI$ & $\mathbb{R}^\numFlux$ & dual variables for flux lower bounds, $l$ \\
    $\dualU$, $\dualUI$ & $\mathbb{R}^\numFlux$ & dual variables for flux upper bounds, $u$ \\
    $\objd$, $\objdI$  & $\mathbb{R}$ & relative importance of the new reaction \\ 
    $\sparse$ & $\mathbb{R}$ & sparsity regularization weight \\
    \midrule
    \multicolumn{3}{c}{\emph{Factor-graph and ADMM variables}}\\
    \midrule
    $\AQP$& \multicolumn{2}{l}{Quadratic function-node}\\
    $\ABI^{(i)}$& \multicolumn{2}{l}{Bi-linear function-node}\\
    $\ABD$& \multicolumn{2}{l}{Boundary function-node}\\
    $\ASP$& \multicolumn{2}{l}{L$1$-norm function-node}\\
    $\AFX$& \multicolumn{2}{l}{Fluxes variable-node}\\
    $\ANR$& \multicolumn{2}{l}{New reaction Variable-node}\\
    $\ADB$& \multicolumn{2}{l}{Dual variable for flux bounds variable-node}\\
    $\ADE^{(i)}$& \multicolumn{2}{l}{Dual variable for equilibrium constraint variable-node}\\
    $\AX$ & \multicolumn{2}{l}{Primal iterate}\\
    $\AZ$& \multicolumn{2}{l}{Consensus iterate}\\
    $\AU$& \multicolumn{2}{l}{Dual iterate}\\
    $\AProx$& \multicolumn{2}{l}{Proximal operator (PO) map}\\ 
    $\Arho$& \multicolumn{2}{l}{ADMM PO tuning parameter}\\
    $\Agama$& \multicolumn{2}{l}{ ADMM over-relaxation tuning parameter}\\
    $\Aalpha$& \multicolumn{2}{l}{ ADMM step-size tuning parameter}\\
    $\An$, $\AnI$& \multicolumn{2}{l}{Input to PO map}\\
    \bottomrule
\end{tabular}
\end{table}

%
%%%%%%%%%%%%%%%%%%%%%%%%%%%%%%%%%
%
\vspace*{-1em}
\section{Solution procedure using ADMM}

BIG-BOSS's inner workings are based on the over-relaxed, distributed consensus, ADMM algorithm \cite{boyd2011distributed,nishihara2015general}. This choice is based on the fact that ADMM (i) allows us to easily parallelize BIG-BOSS, (ii) has good empirical performance for several nonsmooth, nonconvex problems \cite{bento2013message,hao2016fine,derbinsky2013improved,mathysparta,zoran2014shape,derbinsky2013methods}, and (iii) has global optimality guarantees (under mild convexity assumptions) and a convergence rate  that (under proper tuning) equals the convergence rate of the fastest possible first-order method \cite{overrelaxedADMM}, although itself not a first-order method.

Let $f_1(\{\fluxI\}) =     \frac{1}{\numCond}\sum^\numCond_{i=1}
    \norm{\indFI \fluxI-\fluxMesI}^2_2$ and $f_2(\newStoi) = \sparse \| \newStoi \|_1$. Let $f_{\eqref{eq:bilinear_Li_1}}$ be a function that takes the value $+\infty$ if the constraint \eqref{eq:bilinear_Li_1} is not satisfied, and zero if it is. Let $f_{\eqref{eq:bilinear_Li_2}},\dots, f_{\eqref{eq:bilinear_Li_4}}$ be defined analogously.  For the special case of constraints \eqref{eq:bilinear_constraint}, we define one function $f^{(i)}_{\eqref{eq:bilinear_constraint}}$ for each condition $i$ in \eqref{eq:bilinear_constraint}.
To explain how we solve problem \eqref{eq:single_level_inverse_FBA}--\eqref{eq:bilinear_Li_4} using ADMM, it is convenient to perform two simple rerepresentations of \eqref{eq:single_level_inverse_FBA}--\eqref{eq:bilinear_Li_4} using the functions $f_i$. 

\begin{figure}   % Figure 1
\includegraphics[height=6cm, ]{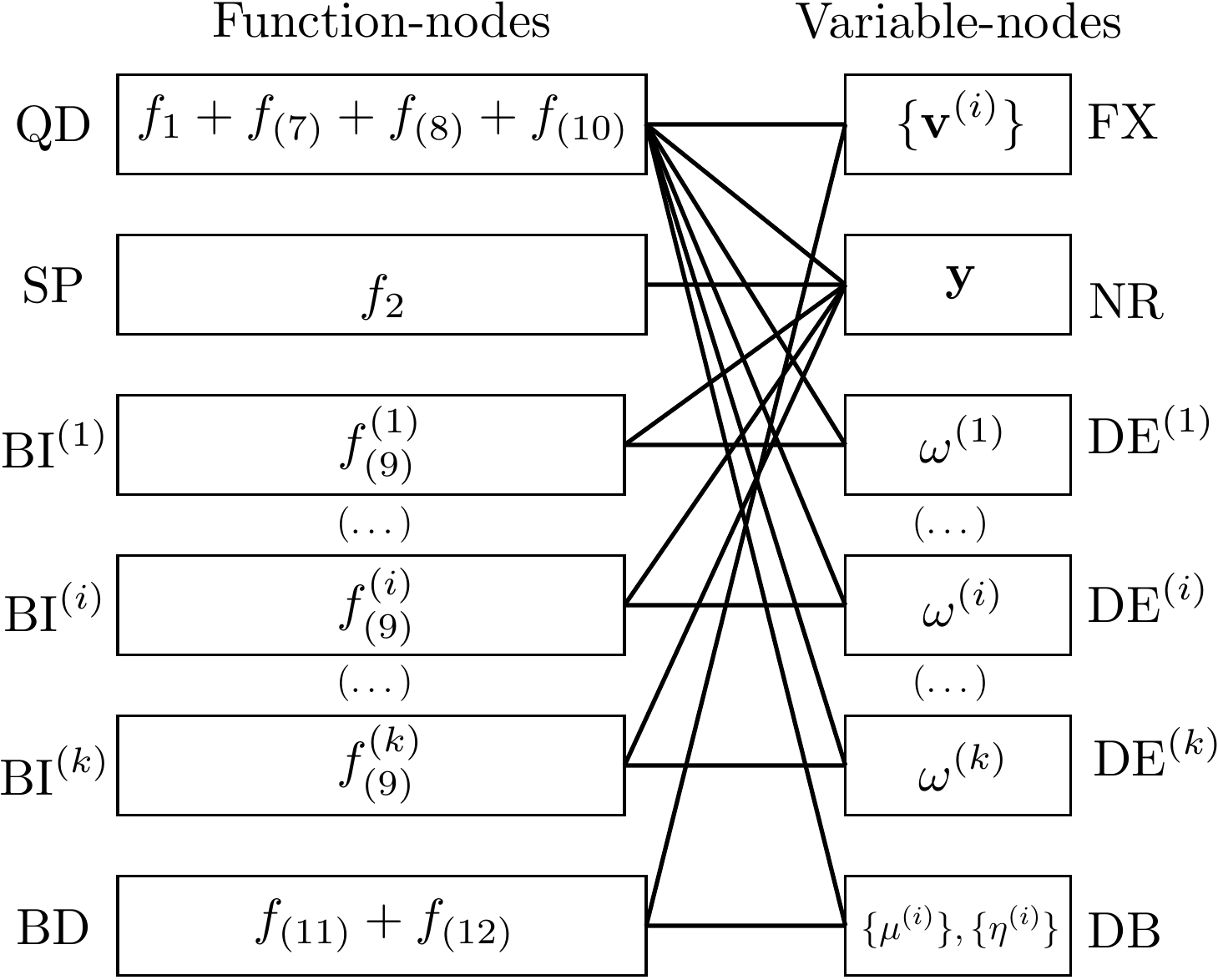}
\caption{Representation of \eqref{eq:single_level_inverse_FBA}-\eqref{eq:bilinear_Li_4} as a factor-graph. Each function-node, and each variable-node, is labeled by two capital letters.}\label{fig:bipartite}
\end{figure}

First, we rewrite problem \eqref{eq:single_level_inverse_FBA}--\eqref{eq:bilinear_Li_4} as $$\min f_1 + f_2 + f_{\eqref{eq:bilinear_Li_1}} + f_{\eqref{eq:bilinear_Li_2}} + f^{(1)}_{\eqref{eq:bilinear_constraint}}+ \dots + f^{(\numCond)}_{\eqref{eq:bilinear_constraint}} + f_{\eqref{eq:bilinear_Li_3}} + f_{\eqref{eq:bilinear_Li_4}}.$$
Second, we represent this unconstrained problem in a \emph{ factor-graph} form, i.e., a bipartite graph connecting objective functions (\emph{function-nodes}) to the variables that they depend on (\emph{variable-nodes}), as shown in Figure \ref{fig:bipartite}. There are $3+\numCond$ function-nodes and variable-nodes in total.

%***************************************
% LY: Function BD should also connect to Variable BD?

We interpret ADMM as an iterative scheme that operates on iterates that live on the edges/nodes of the factor-graph in Figure~\ref{fig:bipartite}, similar to the approaches in \cite{derbinsky2013improved,derbinsky2013methods,hao2016fine}. The function-nodes have the following labels: \emph{quadratic} (\AQP), \emph{sparse} (\ASP), \emph{bi-linear} ($\ABI^{(1)},\dots,\ABI^{(\numCond)}$), and \emph{bound} (BD). The variable-nodes have the following labels: \emph{flux} (\AFX), \emph{new reaction} (\ANR), \emph{dual equilibrium} ($\ADE^{(1)},\dots,\ADE^{(\numCond)}$), and \emph{dual bound} (\ADB).
Each edge (function-node, variable-node), such as $(\ABD,\ADE^{(i)})$, has two associated iterates: a primal iterate $\AX$ and a dual iterate $\AU$. Both iterates are vectors whose dimension matches the variable in the variable-node that they connect to. Each edge also has an associated parameter $\Arho >0$.
For example, the dual iterate and parameter associated with the edge $(\ABD,\ADE^{(i)})$ are $\AU_{\ABD,\ADE^{(i)}}$ (dimension $\numMet$) and $\Arho_{\ABD,\ADE^{(i)}}$ respectively. Each variable-node has one associated iterate: a consensus iterate $\AZ$ whose dimension matches the variable associated with the variable-node. For example, the consensus iterate associated with node $\ADB$ is $\AZ_\ADB$ (dimension $2\numCond\numFlux$).

Essential to ADMM is the concept of a \emph{proximal operator} (PO) of a function $f:\mathbb{R}^s \mapsto \mathbb{R}$ with parameter $\Arho$. This is a map  $\AProx_{f,\Arho}:\mathbb{R}^s \mapsto \mathbb{R}^s$ that takes as input $\An \in \mathbb{R}^s$ and outputs 
\begin{equation}\label{eq:def_of_PO}
\AProx_{f,\Arho}(\An) \in\arg \min_{\dummy \in \mathbb{R}^s} f(\dummy) + \frac{\Arho}{2}\|\An - \dummy\|^2_2.
\end{equation}
 We denote the POs for the function in each function-node as $\AProx_{\AQP}$, $\AProx_{\ASP}$, $\AProx_{\ABI}$, and $\AProx_{\ABD}$ respectively. 
%In this paper, all of the 
All POs have uniquely defined outputs.

Given this setup, the over-relaxed distributed-consensus ADMM algorithm leads directly to Algorithm \ref{alg:admm_big_boss}, which is the basis of BIG-BOSS. In Algorithm \ref{alg:admm_big_boss}, if $Y$ is a variable-node, $\partial Y$ represents all edges incident on $Y$. Similarly, if $X$ is a function-node, $\partial X$ represents all edges incident on $X$.
Note that each for-loop in Algorithm~\ref{alg:admm_big_boss} can be computed in parallel.
\vspace*{-\topsep}
\begin{algorithm}[!ht]
\caption{Algorithm for BIG-BOSS\label{alg:admm_big_boss}}
\begin{algorithmic}[1]
%\WHILE[Solve proximal operators]{$!$\text{stopping criteria}}
\While{$!$\text{stopping criteria}}
    % PO: constrained least-squares
    \For{$E=(X,Y) \in \text{edges of factor-graph}$}
    \State $\AU_E \gets \AU_E + \Aalpha(\Agama\AX_E - \AZ_Y + (1-\Agama)\AZ_Y^\text{old})$
    \EndFor
    \For{$X \in \text{function-nodes of factor-graph}$}
    \State $\{\AX_E\}_{E \in \partial X} \gets \AProx_{X,\Arho_X}(\{\AZ_Y - \AU_E\}_{E \in \partial X})$
    \EndFor
       \For{$Y \in \text{variable-nodes of factor-graph}$}
    \State $\AZ_Y^\text{old} \gets \AZ_Y$
    \State $\AZ_Y \gets (1- \Agama)\AZ_Y + \frac{\sum_{E \in \partial Y} \Arho_E(\Agama \AX_E + \AU_E)}{\sum_{E \in \partial Y} \Arho_E}$
    \EndFor
\EndWhile
\end{algorithmic}
\end{algorithm}
The rate of  convergence of ADMM is influenced by the multiple $\Arho$'s, $\Agama$, and $\Aalpha$, and there are various methods for choosing
good values or adapting them while ADMM is running 
\cite{he2000alternating, boyd2011distributed,
derbinsky2013improved,
xu2016adaptive, xu2017admm, osqp, wohlberg2017admm}.
As described in Section \ref{section:adaptive}, we
implement the residual balancing method \cite{osqp}. Although this scheme is not provably optimal, the algorithm converges very quickly in practice, and the BIG-BOSS user need not worry about tuning in general.
In BIG-BOSS, all of the $\Arho$'s corresponding to edges incident on the same function-node $X$ in the factor-graph have the same value, $\Arho_X$, throughout ADMM's execution. See Section \ref{section:adaptive} for more details.
In this work, we run ADMM until all of the iterates
are identical between iterations to a numerical absolute tolerance
of $10^{-9}$. This is our \emph{stopping criterion}.

Most updates in Algorithm \ref{alg:admm_big_boss} are linear. 
The exceptions are updates involving POs, %i.e., $\AProx$, 
which require solving  \eqref{eq:def_of_PO} for different functions $f$. A vital part of our work was to find efficient solvers for each PO. %In fact, o
Our decomposition of problem \eqref{eq:single_level_inverse_FBA}-\eqref{eq:bilinear_Li_4} into the specific form of Figure \ref{fig:bipartite} is not accidental. It leads to a very small number of POs, all of which can be evaluated quickly. ($\AProx_{\AQP}$ amounts to solving a linear system, and the others can be computed in closed form.)
We expand on this in the next sections.

\vspace*{-1em}
%------------------------------------------------------------
%\subsection{Proximal operator for least-squares minimization}
\subsection{Proximal operator for node $\AQP$}
%------------------------------------------------------------
%The proximal operator $\mathbf{Prox}_{\text{LS}}(\cdot)$ 
%minimizes the least-squares error subject to linear constraints.
The function-node $\AQP$ is a quadratic objective subject to some linear constraints. To explain how we compute $\AProx_{\AQP}$, we start by defining $\dummyI = \{ \fluxI, \dualSI, \dualLI, \dualUI, \newStoi \} \in \mathbb{R}^{2(\numFlux+\numMet)+\numMet}$ for each condition $i$, writing
$$f_1(\{\dummyI\}) = \frac{1}{k}\sum^{\numCond}_{i =1}  \norm{ \blkFI \dummyI - \blkbI}_2^2,$$
and writing the constraints $f_{\eqref{eq:bilinear_Li_1}} + f_{\eqref{eq:bilinear_Li_2}} + f_{\eqref{eq:dual_equals_primal}} < +\infty$ as 
 $\blkLhsI \dummyI = \blkRhsI$, $\forall i \in \{1,\dots,\numCond\}$, where 
$$
\blkFI =
\begin{bmatrix}
    \indFI & 0 & 0 & 0 & 0
\end{bmatrix},\ 
\blkbI =
\begin{bmatrix}
    \fluxMesI 
\end{bmatrix},
$$
$$
\blkLhsI =
\begin{bmatrix}
    \stoi & 0 & 0 & 0 & \newFluxI \\
      0   & \stoi\T & -I & I & 0\\
    \objc\T & -\bal\T & \lowBI\T & -\uppBI\T & 0
\end{bmatrix}, \text{ and }
\blkRhsI = 
\begin{bmatrix}
    \balI \\
    \objc \\
    -\newFluxI
\end{bmatrix}.
$$
Evaluating $\AProx_{\AQP}(\{\AnI\})$ for input vectors $\{\AnI\} \in \mathbb{R}^{2\numCond(\numFlux+\numMet)+\numMet}$ now amounts to finding the unique minimizer of a strongly convex quadratic program with linear constraints, namely, 
\begin{align*}
&\min_{\{\dummyI\} \in \mathbb{R}^{2k(\numFlux+\numMet)+\numMet}} \frac{1}{k}\sum^{\numCond}_{i=1}\|\blkFI \dummyI - \blkbI\|^2_2 +  \frac{\Arho}{2} \sum^{\numCond}_{i=1}  \|\dummyI - \AnI\|^2_2\\
&\text{ subject to } \blkLhsI \dummyI = \blkRhsI, \forall i \in \{1,\dots,\numCond\}.
\end{align*}
We find this unique minimizer by solving a linear system 
obtained from the KKT conditions. 
We then have $\numCond$ linear systems that are coupled by
the variable $\newStoi$ that is common to each $\dummyI$.
%This linear system 
%decomposes 
%into $\numCond$ 
%%independent 
%linear systems, one per condition $i$,
We can write each linear system in block form as 
\begin{equation}
\begin{bmatrix}
    \frac{2}{k}\blkFI\T \blkFI + \Arho I   & \blkLhsI^T \\
    \blkLhsI               & 0
\end{bmatrix}
\begin{bmatrix}
    \dummyI \\
    \dualdummyI
\end{bmatrix}
= 
\begin{bmatrix}
    \frac{2}{k}\blkFI\T \blkbI + \Arho \AnI  \\
    \blkRhsI
\end{bmatrix},
\label{eq:leastsq_kkt}
\end{equation}
where $\dualdummyI$ is the Lagrange multiplier
for the linear constraints $\blkLhsI \dummyI = \blkRhsI$.
Finally, we stack the $\numCond$ linear systems into
one large linear system\footnote{If $\numCond$ is very large,
we can also split the function-node $\AQP$ into $\numCond$ function-nodes, one per condition $i$. This will result in $\numCond$ smaller linear systems, now decoupled, that can be solved in parallel.}.
%, even distributing them across multiple machines if necessary.
%
%%% To ensure matrix is nonsingular
For numerical stability reasons, 
%we change the zero block to
we add $-\beta I$, $\beta=10^{-12}$, to the system's matrix.
This ensures a unique solution.
As described in Section~\ref{sec:precond}, we
also implemented a preconditioner for this linear system to improve its condition number.

To solve the linear system, we compared two different numerical libraries. One is the C library
UMFPACK, which we called from Fortran using the 
Fortran 2003 interface in mUMFPACK 1.0 \cite{mUMFPACK}.
The other is the symmetric indefinite solver
MA57 \cite{duff2004ma57} from HSL \cite{hsl2013}.
For both solvers, the symbolic factorization needs to be
computed only once. When $\rho$ is updated,
only the numerical factorization is recomputed.
As other groups have reported \cite{gould2007numerical},
we found that MA57's solve phase was faster than UMFPACK's.
We therefore used MA57 for all results in this study.

\subsection{Proximal operator for node $\ADB$}

Observe that the constraint $f_{\eqref{eq:bilinear_Li_3}}+f_{\eqref{eq:bilinear_Li_4}} < +\infty$ is equivalent to $3\numCond$ constraints of the type $\blkRhs^{(i,j)} \leq \dummy^{(i,j)} \leq \blkb^{(i,j)}$ for some constant vectors $\blkRhs^{(i,j)},\blkb^{(i,j)} \in  \mathbb{R}^{\numFlux}$: one for each condition $i$ and  each of the variables $\fluxI$ (for index $j=1$), $\dualLI$ ( for index $j=3$), and $\dualUI$ (for index $j = 4$). 
Hence, computing $\AProx_{\ADB}(\{\An^{(i,j)}\})$ for some input $\{\An^{(i,j)}\} \in \mathbb{R}^{3\numCond\numFlux}$ amounts to solving $3\numCond$ independent problems 
$$
\min_{\dummy^{(i,j)} \in \mathbb{R}^{\numFlux}} \frac{\rho}{2} \norm{ \dummy^{(i,j)} - \An^{(i,j)}}_2^2
\text{\ subject to\ } \blkRhs^{(i,j)} \leq \dummy^{(i,j)} \leq \blkb^{(i,j)}.
$$
This problem has a unique minimizer given explicitly by
$$
\dummy^{(i,j)} = \min\{\blkb^{(i,j)},\max\{\blkRhs^{(i,j)},\An^{(i,j)}\}\},
$$
where $\min$ and $\max$ are taken component-wise.

%------------------------------------------------------
\subsection{Proximal operator for node $\ASP$}
%------------------------------------------------------

The PO for node SP, which has $f_2(\dummy) = \delta\|\dummy\|_1$, requires finding the minimizer of 
% $$
% \argmin_{x\in \mathbb{R}^n} \frac{\rho}{2}\|x-N^k \|_2^2 +
% \sparse |x|,
% $$
$$
\min_{\dummy \in \mathbb{R}^{\numMet} } \frac{\rho}{2} \norm{ \dummy - \An}_2^2 
+ \sparse \norm{\dummy}_1.
$$
This minimizer is
% $$
% x_{L1}^{k+1} = \mathbf{S}_{\sparse/\rho}(x - N^k),
% $$
$
\dummy = \mathbf{S}_{\sparse/\rho}(\An)$, 
where $\mathbf{S}_{t}(\cdot)$ is the element-wise
soft-thresholding operator \cite{boyd2011distributed}
defined, for input scalar $x$, as
\begin{equation}
S_t(x) = 
\begin{cases}
    0,  & \text{if}\  |x| \leq t \\
    \text{sign}(x)(|x| - t), & \text{otherwise}.
\end{cases}
\end{equation}
%

%------------------------------------------------------
\subsection{Proximal operator for node $\ABI^{(i)}$}
%------------------------------------------------------

%We define here the bilinear proximal operator,
%$\mathbf{Prox}_{\text{Bi}}(n)$.
% For convenience of notation, we will 
% split $x$ into equal length
% vectors $x$ and $y$, and define the 
% constraint $y^T x \geq d$ \eqref{eq:bilinear2}.
%***************************************************
The $i$th bilinear proximal operator $\AProx_{\ABI^{(i)}}$ receives input vectors
$\An, \AnI \in \mathbb{R}^{\numMet}$, and outputs the minimizer of
\begin{align}
\min_{\newStoi , \dummyI \in \mathbb{R}^{\numMet}}
\quad & \frac{\rho}{2} \norm{ \newStoi - \An }^2_2 + 
     \frac{\rho}{2} \norm{ \dualSI - \AnI }^2_2 \label{eq:bilinear1}
        \\
\text{subject to} \quad & \newStoi\T \dualSI \geq \objdI.
\label{eq:bilinear2}
\end{align}
%
% The closed form solution to this problem is the following: 
% \begin{equation}
% \{\hat{\newStoi}, \hat{\dualS}^{(i)} \} = 
% \begin{cases}
%     \{\nnewStoi, \ndualSI\},  & \nnewStoi \ndualSI \geq \objdI \\
%     , & \text{otherwise}.
% \end{cases}
% \end{equation}
%
To solve this problem, we consider two cases. If $\AnI\T \An \geq \objdI$, we know that the minimizer is $(\newStoi , \dummyI) = (\An, \AnI)$. Otherwise, we know that the constraint \eqref{eq:bilinear2} is active, and we compute the minimizer from among the set of solutions of the KKT conditions for the (now constrained) problem. The KKT conditions are 
\begin{align*}
    \rho \newStoi - \rho \An -\dualdummy \dualSI  = 0, \  
    \rho \dualSI - \rho \AnI -\dualdummy \newStoi = 0, \ 
    \newStoi\T \dualSI = \objdI,
\end{align*}
where $\dualdummy$ is the dual variable for the constraint $\newStoi\T \dualSI \geq \objdI$.
%
% \begin{align}
% \nabla_x L(x,y,\lambda) &= \rho x - \rho N_x -\lambda y = 0 \\
% \nabla_y L(x,y,\lambda) &= \rho y - \rho N_y -\lambda x = 0 \\
% y^T x &\geq d \\
% \lambda (d  - y^T x) &= 0.
% \end{align}
%
By defining $x = \dualdummy / \Arho$ and reformulating the KKT system, we arrive at the following 
quartic equation that determines the value of $\dualdummy$:
\begin{align}
    &\objdI x^4 + (-2 \objdI - \An\T \AnI) x^2 + (\An\T \An+ \AnI\T \AnI) x \nonumber
\\  &\null +  (\objdI - \An\T \AnI) = 0.
\end{align}
% $$
% d z^4 + (-2 d - N_x^T N_y) z^2 + (N_x^T N_x + N_y^T N_y) z + (d-N_x^T N_y) = 0.
% $$
%
%The roots of this equation are readily found as a closed form solution or using standard root solvers.
%
We solve this quartic equation in closed form,
using the Quartic Polynomial 
Program Fortran code \cite{morris1993nswc,pdas}.
For each real root, $x$, we compute $\newStoi$ and $\dualSI$ as
\begin{align}
\newStoi &= \frac{\An - {x} \AnI}{
    1 - {x}^2 },
    \quad
\dualSI = \frac{ \AnI - {x} \An}{
    1 - {x}^2 },
\end{align}
and we keep the solution that minimizes the objective 
function \eqref{eq:bilinear1}.
%From this root, we recover $\lambda = \rho x$.

%
% \begin{align}
% \newStoi &= \frac{\rho \nnewStoi - \lambda \ndualSI}{
%     \rho - \lambda^2 / \rho }, \\
% \dualSI &= \frac{\rho \ndualSI - \lambda \nnewStoi}{
%     \rho - \lambda^2 / \rho }.
% \end{align}
%
% \begin{align}
% x &= \frac{\rho N_x - \lambda N_y}{ \rho - \lambda^2 / \rho }, \\
% y &= \frac{\rho N_y - \lambda N_x}{ \rho - \lambda^2 / \rho }.
% \end{align}
%

% Finally, we note that general quadratic constraints 
% were handled in a similar manner by
% \cite{huang2016consensus}

%-------------------------------------------
\subsection{Preconditioner} \label{sec:precond}
%-------------------------------------------
The convergence rate of the ADMM can be improved by preconditioning
the data matrices associated with the node $\AQP$ \cite{osqp,takapoui2016preconditioning}.
%by applying a diagonal scaling matrix to
%the problem 
%$$
%\min \ f(x) + g(z) \ \text{subject to}\ x = z
%$$
%to yield the scaled problem,
%$$
%\min \ f(x) + g(z) \ \text{subject to}\ Fx = Fz
%$$
%preconditioning to a 
%$L_1$-regularized lasso problem, only X iterations were
%needed, compared with over 7,400 ADMM iterations 
%without any preconditioning
%\cite{lasso_preconditioner}.
%

%\begin{align}
%    \min_x \quad & \|Fx - b\|_2^2 + \delta |Lx| \\
%    \text{subject to} \quad & Cx = d \\
%                            & l \leq x \leq u,
%\end{align}
We use the Ruiz equilibration method \cite{ruiz2001scaling}
to compute 
positive definite diagonal matrices, 
$\scaleUL$ and $\scaleLR$, with which we scale all of the variables $\{\dummyI= \{\fluxI, \dualSI ,\dualLI,\dualUI,\newStoi\}\}$ in our formulation, as well as the constraints 
$\blkLhsI \dummyI = \blkRhsI \forall i$, associated with the node $\AQP$.
The new scaled variables are defined through the change of variable $\dummyI  \rightarrow \scaleUL {\dummyI}$, and the new constraints then transform as $ \blkLhsI  {\dummyI} = 
\blkRhsI \rightarrow \scaleLR \blkLhsI \scaleUL {\dummyI} = \scaleLR \blkRhsI \forall i$. Note that all of the copies of $\newStoi$ inside each $\dummyI$ get scaled exactly in the same way.

This scaling affects the different proximal operators in different ways. The node $\AQP$, that requires computing   $\AProx_{\AQP}(\{\AnI\})$, now requires solving a modified version of \eqref{eq:leastsq_kkt}, namely, the following system of linear equations (coupled by $\newStoi$),
\begin{equation*}
\begin{bmatrix}
    \scaleUL \frac{2}{k}\blkFI\T \blkFI \scaleUL + \Arho I  &
        \scaleUL \blkLhsI\T \scaleLR \\
    \scaleLR \blkLhsI \scaleUL & 0
    %DPD & DC^T E \\
    %ECD & 0
\end{bmatrix}
\begin{bmatrix}
    {\dummyI} \\
    {\dualdummyI}
\end{bmatrix}
=
\begin{bmatrix}
    \scaleUL \frac{2}{k}\blkFI\T \blkbI + \Arho {\AnI} \\
    \scaleLR \blkRhsI 
    % D q \\
    % E d 
\end{bmatrix}.
\end{equation*}
%

%%% TODO
%%% And lambda not seen 
%%% Communicates scaled variables to other proximal operators.
%We used the Ruiz equilibration method \cite{ruiz2001scaling}
%to compute $\scaleMat$.
%This method reduced the condition number
%of the left-hand-side matrix of \eqref{eq:leastsq_kkt} by 
%approximately three orders of magnitude.

%For the bound proximal operator, the scaled solution is
%For node $\ADB$, the scaled solution is
For $\AProx_{\ADB}(\{\An^{(i,j)}\})$, let $\scaleUL^{(j)}$, $j=1,3,4$, be the diagonal block of $\scaleUL$ that  scales the variables $\fluxI$ ($j=1$), $\dualLI$ ($j=3$), and $\dualUI$ ($j=4$).  We now solve $3k$ problems of the type
\begin{align}
&\min_{\dummy^{(i,j)} \in \mathbb{R}^{\numFlux}} \frac{\rho}{2} \norm{ {\dummy^{(i,j)}} - {\An^{(i,j)}}}_2^2
\text{\ subject to\ }\\ &{\scaleUL^{(j)}}^{-1} \blkRhs^{(i,j)} \leq {\dummy^{(i,j)}} \leq {\scaleUL^{(j)}}^{-1} \blkb^{(i,j)},
\end{align}
%\scaled{\dummy} = \Pi_{\scaled{B}}(\dummySc - \ndummySc),
which has a unique minimizer
$$
{\dummyI} = \min\{{\scaleUL^{(j)}}^{-1} \blkb^{(i,j)}, \max\{{\scaleUL^{(j)}}^{-1} \blkRhs^{(i,j)}, {\An^{(i,j)}}\}\}.
$$
%$$
% $$
% \tilde{x}^{k+1} = \Pi_{\tilde{B}}(\tilde{x} - n^k),
% $$
%where $\Pi_{\scaled{B}}$ projects $\dummySc-\ndummySc$
%onto the domain $\{ \scaleUL^{-1}\lowB, \scaleUL^{-1}\uppB\}$.
% where $\Pi_{\tilde{B}}$ projects $\tilde{x}-n^k$
% onto the domain $\{D^{-1}l, D^{-1}u\}$.

%For the $L_1$ %5regularization
%proximal operator, 
For the node $\ASP$, let $\scaleUL^{(5)}$ be the diagonal block of $\scaleUL$ that scales the variable $\newStoi$. The new $\AProx_{\ASP}(\An)$ now outputs
$$
{\dummy} = 
\mathbf{S}_{{\scaleUL^{(5)}}^{-1} \sparse / \rho} (\An),
$$
where, for the $r$th component of $\An$, we use the $r$th diagonal element of  ${\scaleUL^{(5)}}^{-1} \sparse / \rho$ in  the operator $\mathbf{S}$, that is applied component-wise.
% $$
% \tilde{x}^{k+1} = \mathbf{S}_{D^{-1} \sparse / \rho} (x - n^k).
% $$

%For the bilinear proximal operator, the scaled solution is
For the $i$th bilinear proximal operator $\AProx_{\ABI^{(i)}}(\An, \AnI)$, 
we now solve
\begin{align*}
\argmin_{{\newStoi},{\dualSI} \in \mathbb{R}^{\numMet}} \quad & 
    \frac{\rho}{2} \norm{ {\newStoi} - {\An} }^2_2 +  \frac{\rho}{2} \norm{ {\dualSI} - {\AnI}^{} }^2_2 \\
\text{subject to} \quad & {\newStoi}\T \scaleUL^{(5)} \scaleUL^{(2)} {\dualSI}
\geq \objdI,
\end{align*}
where $\scaleUL^{(2)}$ is the diagonal block of $\scaleUL$ that scales the variable $\dualSI$. We modify the Ruiz equilibration method such that $\scaleUL^{(5)} \scaleUL^{(2)} = constant\times I$, and hence we can solve this PO using the same method as before.

\subsection{Adaptive penalty} \label{section:adaptive}
%--------------------------------------------------

%%%% LY: Need to update so clear every PO's rho is adapted.

Our ADMM scheme allows for one $\Arho$ per each edge in the factor-graph in Fig. \ref{fig:bipartite}. In BIG-BOSS, these $\Arho$'s are chosen automatically, such that a user does not need to worry about tuning. 
All of the $\Arho$'s associated with edges incident on the same function-node have the same value. Hence, we refer to the different $\Arho$'s by their respective PO.
%
% The $\Arho$'s for $\AQP$, which is the only PO that depends on the input data, are chosen dynamically, to speedup convergence.
%
We update the $\Arho$'s for each PO dynamically every $1000$ iterations using the \emph{residual balancing method} \cite{boyd2011distributed}, which we now explain.

Similar to \cite{osqp,wohlberg2017admm}, we define the vector of
relative primal residuals at iteration $10^3 \times t$, and for any proximal operator $X$, as
\[
r_{\text{prim}}(t) =
\frac{ \{\AX_E(t) - \AZ_Y(t)\}_{E: E = (X,Y)\in \partial X} }{
     \max \{\{\max\{|\AX_E(t)|, |\AZ_Y(t)|\}\}_{E: E = (X,Y)\in \partial X}\}},
\]
where, we recall, $\partial X$ are all the edges incident on node $X$.
%
% \[
% r_{\text{prim}} =
% \frac{ \| \epsilon_\text{primal} \|_{\infty} }{
%     \max\{ \|A x^k\|_{\infty}, \|z^k\|_{\infty} \}
% }
% \]
%

The vector of relative dual residuals for the proximal operator $X=\AQP$ at iteration $10^3 \times t$ is
defined as follows. Let 
\begin{align*}
    C_1 &= \max^\numCond_{i = 1}  \left\| 
    \frac{2}{k}\blkFI\T\blkFI \{\AX_E(t+1)\}_{E\in\partial X} \right\|_\infty ,\\
    C_2 &=  \| \{\AU_E(t+1)\}_{E\in \partial X} \|_\infty, \text{ and } C_3 = \max^{\numCond}_{i=1}  \left \| -\frac{2}{k}\blkFI\T\blkbI \right \|_\infty .
\end{align*}
We define
$$r_{\text{dual}}(t+1) = \frac{\{\AZ_Y(t+1) - \AZ_Y(t)\}_{Y:(X,Y)\in \partial X}}{\max\{C_1,C_2,C_3\}}.$$

If $X$ is a PO other than $\AQP$, we define
$$r_{\text{dual}}(t+1) = \frac{\{\AZ_Y(t+1) - \AZ_Y(t)\}_{Y:(X,Y)\in \partial X}}{C_2},$$
since $C_1$ and $C_3$ do not exist when $X$ is not $\AQP$.

The $\Arho$ update rule is then defined as
follows. Once the $\infty$-norm of both the primal and the dual residuals falls below
a threshold of $10^{-3}$, we start updating $\Arho$ as
$$\Arho(t+1) = \max\{10^{-3},\min\{10^{3},\tilde{\Arho}(t+1)\}\}$$ where
\begin{equation*}
\tilde{\Arho}(t+1) =
\begin{cases}
    \tau \Arho(t)    & \text{if}\
        \left|\ln\left(\|r_{\text{prim}}(t)\|_\infty / 
        \|r_{\text{dual}}(t)\|_\infty  \right) \right| \geq \ln(\eta), \\ 
    \rho(t)                 & \text{otherwise},
\end{cases}
\end{equation*}
where $\tau = \sqrt{\|r(t)_\text{prim}\|_\infty / 
                    \|r(t)_\text{dual}\|_\infty }$
and $\eta>0$, which in this study we set to $2$, is a threshold to prevent the $\Arho$
from changing too frequently. 
Note that a frequently-changing  $\Arho$ requires many numerical re-factorization for solving the 
linear system in the node $\AQP$. Furthermore, a frequently-changing $\Arho$ might compromise the convergence of the ADMM.
%

%

%
%
%\[
%r_{\text{dual}} =
%\frac{\| \epsilon_\text{dual} \|_{\infty} }{
%    \max\{ \| P x^k \|_{\infty}, \| A^T y^k\|_{\infty} , \| q \|_{\infty}\} 
%}
%\]
%

%
% \[
% \epsilon_{\text{rel}} =
% \sqrt{
% \frac{
%     \| \epsilon_\text{primal} \|_{\infty} / \max\{
%         \|A x^k\|_{\infty}, \|z^k\|_{\infty}
%     \}
%     }{
%     \| \epsilon_\text{dual} \|_{\infty} / \max\{
%     \| P x^k \|_{\infty}, \| A^T y^k\|_{\infty} , \| q \|_{\infty}\} 
%     }
% }
% \]
%

%--------------------------------------------------
\subsection{Step size} \label{section:stepsize}
%--------------------------------------------------
The step size $\stepsize$ in Algorithm~\ref{alg:admm_big_boss} can be
adjusted to speed up convergence.
We used a default step size of 1.0 for all problems.
If the problem did not converge to the desired tolerance
within 2 million iterations, we re-ran the problem
with a simple step size heuristic.
We use $\stepsize=1$ for 10,000 iterations,
$\stepsize=0.1$ for 100,000 iterations, then
$\stepsize=0.001$ until convergence, or the 
iteration limit is reached.

%==================================================
\section{Experimental results}
\subsection{BOSS versus BIG-BOSS \label{sec:compareboss}}
%--------------------------------------------------

We compare \ouralg{} with BOSS in terms of running speed, and accuracy.
To make a valid comparison, we set $\numCond=1$ and $\sparse=0$, such that \ouralg{} is doing the same inference as BOSS  \cite{boss2008}.
BOSS requires choosing an off-the-shelf nonlinear programming solver, we choose  %solve this problem using
IPOPT \cite{wachter2006ipopt} (v24.7.1),
an open-source, interior-point solver used to solve
large-scale nonlinear programs.
%with Intel MKL PARDISO as the linear solver.

We generate synthetic flux measurements, $\fluxMes$,
using the latest metabolic network of 
\textit{E. coli} \cite{iML1515} called iML1515. This network has $1,877$ metabolites and $2,712$ reactions. We then hide one goal reaction from the
network, $\newStoi$, and try to infer it from flux data.

Using a single CPU core (Intel i7, $3.4$GHz), BOSS is able to learn $\newStoi$ for this model in $13.3$ seconds (wall time) using IPOPT to primal and dual residuals of
$1.05\times 10^{-11}$ and $7.23\times 10^{-11}$. Using the same CPU core,
\ouralg{} is able to learn $\newStoi$ in $113$ seconds to 
primal and dual residuals of $2.73\times 10^{-10}$
and $6.63\times 10^{-10}$. In Appendix \ref{apps:other_acc}, 
we report the run-time for other accuracy values. 
Although \ouralg{} is slower than BOSS, it is still fast, and, 
more importantly, allows $k > 1, \delta > 0$, and the use of multiple cores for large models.

Now we look at how well
the recovered reaction coefficients $\newStoi$ compare to the true coefficients that we hide, both in BOSS and \ouralg{}. We do so by (i) directly comparing the ground truth with the recovered $\newStoi$, as well as, (ii) using the recovered $\newStoi$ to predict measured fluxes (training fluxes).

\begin{figure}[!ht]
\centering
\includegraphics[width=1.1\columnwidth]{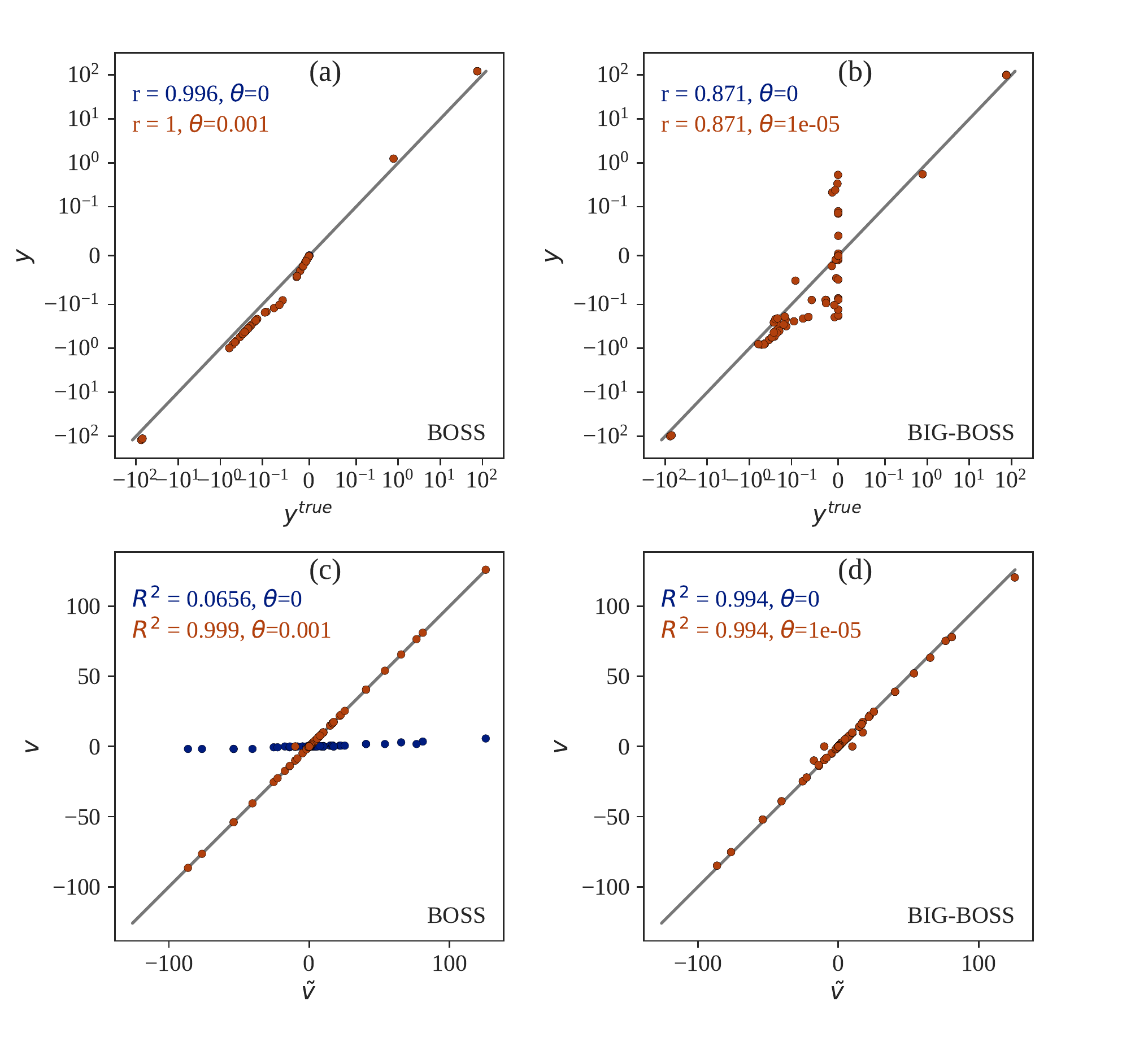}
\caption{Performance of BOSS and \ouralg{}.
    Coefficients of $\newStoi$ estimated using BOSS (a)
    and \ouralg{} (b) compared against true coefficients.
    Fluxes predicted using the goal reaction estimated
    by BOSS (c) and \ouralg{} (d)
    compared against training fluxes $\fluxMes$.
    $R^2$: coefficient of variation. $r$: Spearman
    rank correlation. $\theta$: threshold for zeroing out
    elements of $\newStoi$ smaller than $\theta$.
    \label{fig:compareboss}
}
\end{figure}

For BOSS, the recovered goal reaction coefficients $\newStoi$ closely resemble the true 
reaction coefficients, with Spearman rank correlation, 
$r^{\text{Spearman}} = 0.996$ 
(Fig.~\ref{fig:compareboss}a).
With the recovered $\newStoi$ back into the model, we then solve \eqref{eq:FBA} to simulate fluxes.
%***********************************************************
%
For BOSS, the estimated goal reaction has zero flux,
resulting in simulated fluxes that do not resemble
the training fluxes (Fig.~\ref{fig:compareboss}c).
To help BOSS, we 
make $\newStoi$ sparse by zeroing coefficients
with absolute value smaller than a fixed threshold $\theta$.
Since this cannot be done in a real setting, it shows a limitation in BOSS.
We tested thresholds between $0$ to $1$
and found that values between
$10^{-3}$ to $10^{-1}$ result in 
accurate flux predictions ($0.98\leq R^2 \leq 1.0$).
%However, the true reaction has
%$70$ non-zero coefficients with magnitudes between
%$2\times 10^{-6}$ to $75.6$. 

%%%% Test performance on test conditions 

%***********************************************************

We perform the same experiment using \ouralg{}, which estimates 
$\newStoi$ with 
$r^{\text{Spearman}} = 0.871$ (Fig.~\ref{fig:compareboss}b).
Fluxes predicted using this new reaction are accurate on the training data 
to $R^2 = 0.994$ without zeroing additional coefficients
(Fig.~\ref{fig:compareboss}d).
%%%
%%% Should get better R^2 with smaller delta
%%%

% BOSS Convergence: 
% Dual infeas (unscaled): 7.23 e-11
% Constraint viol (unscaled): 1.05E-11
% Obj (unscaled): 3.61E-7

%--------------------------------------------------
%--------------------------------------------------
%--------------------------------------------------
%\subsection{Effect of missing measurements}
\subsection{Experiments with $75$ large-scale models}
%--------------------------------------------------
% \subsubsection{Metabolic reaction fluxes}
% Current 13-C labeled metabolic flux analysis (MFA) technology 
% is capable of measuring a subset of reaction fluxes.
% The number of measured fluxes is on the order of dozens \cite{MFA}.
% More recent technologies have measured up to X reactions \cite{gsMFA}.
% Thus, for a model having over 2000 reactions, approximately $x-y$\%
%is measured.
%While relatively few fluxes are measured, they are informative.
%Therefore, 
%Therefore, in addition to testing random measurements, we also 
%considered a nominal reaction set based on
%typically measured reactions. We then subtracted or added measured
%reactions from this baseline set.

We test our algorithm's performance on $75$ large-scale 
metabolic networks in the BiGG database \cite{bigg_models}. 
Model sizes ranged from $95$ to $10,600$ reactions
and $72$ to $5,835$ metabolites. The median model size
is $2,727$ reactions and $1,942$ metabolites.
For each model, we generate synthetic training fluxes
by solving Problem~\eqref{eq:FBA} using the existing
cellular goal reaction. 
To generate test data, we do as above but in a different
simulated nutrient environment. This is accomplished by
changing the bounds $\lowB$ and $\uppB$ on
nutrient transport reactions.
Our algorithm then receives as input
the training fluxes, the bounds, and an incomplete stoichiometric
matrix, which is missing the column corresponding to the 
``ground truth'' cellular goal reaction that we want to estimate.

We evaluate our algorithm's performance based on three
metrics: (i) how well the estimated reaction allows to predict training fluxes, (ii) how well the estimated reaction allows to predict testing fluxes, and
(iii) how close the estimated reaction coefficients are to the ground truth coefficients.
To evaluate metrics (i) and (ii), 
we insert the estimated reaction
into the metabolic network, having first removed 
the true goal reaction, and then simulate fluxes.

Based on the coefficient of determination ($R^2$) 
between measured and predicted fluxes,
both training and test performances were 
high (median $R^2$ of $0.993$ and $0.977$) (Table~\ref{tab:bigg_flux}).
To assess the significance of these $R^2$ values,
we compare them with fluxes generated
from $1,000$ random goal reactions. 
These random goal reactions are obtained by 
randomly shuffling the coefficients of the ground truth coefficients.
We say that the $R^2$ based on our algorithm is significant if it exceeds 
95\% of the $R^2$ values that were computed using the 1,000 random goal reactions.
Overall, $R^2$ values are significant for $70.1\%$ of the models
for training data and $67.2\%$ of the models for test data
(Table~\ref{tab:bigg_flux}).
%(Fig.~\ref{fig:perf_signif}).
%
%We assessed the accuracy of recovering the ground truth goal reaction
The ground truth goal reaction is recovered accurately:
across $75$ models, the median Pearson correlation was $1.0$
(Table~\ref{tab:objrec}). 
The Spearman rank correlation was lower, with median $0.48$.
%The median absolute percent error (MdAPE) of all reaction
%coefficients was also good, with median value 0\%.
The fact that predicted fluxes are accurate indicates
that metabolic flux predictions are relatively 
insensitive to the precise recovery of the true coefficients in $\newStoi$.

In biological experiments, the majority of fluxes
are unmeasured, so $50\%$ missing flux measurements is 
common \cite{antoniewicz2015methods}.
We test the effect of missing flux measurements
by randomly removing $10$\% and $50$\% of measurements,
repeated five times.
The main consequence of missing measurements is
that recovering the ground truth goal is
more difficult.
With $10\%$ and $50\%$ of missing measurements,
median Pearson correlations of $-0.023$ and $0.0012$ are
obtained (Table \ref{tab:bigg_flux}).
The accuracy of predicted fluxes also
deteriorate when measurements are missing 
(Table \ref{tab:bigg_flux}).
However, depending on which reactions
are measured, fluxes can be predicted with 
$R^2 > 0.90$ for $6.67\%$ of the cases with $10\%$ of missing measurements,
and for $3.16\%$ of the cases with $50\%$ of missing measurements, in the case of test data.
This result shows that certain fluxes are more 
important than others,
and that \ouralg{} can recover goal reactions
when those important fluxes are measured.
%These recovered goals are thus useful for 
%predicting metabolic behavior in certain new conditions.
%This result indicates that even when
%half the flux measurements are missing, a subset of fluxes
%are informative for 
%recovering the goal reaction.
%This result indicates that while recovering
%goal reactions is difficult when measurements are missing,
%a subset of 
%
%Next, we show one way to address this real-world challenge.
%\setlength{\abovecaptionskip}{6pt}
\begin{table}[ht]   % Table 2
\caption{Performance for $75$ BiGG models.
$R^2$: coefficient of determination.
By `significant' we mean that the $R^2$ based on our algorithm's goal reaction 
exceeds 95\% of the $R^2$ values that are based on 1,000 random goal reactions.
\label{tab:bigg_flux}}
\begin{tabular}{lrrrrr}
\toprule
     &  \% missing &   \multicolumn{2}{c}{$R^2$} & \% cases, &\% cases,\\ 
     \cline{3-4}
Data & fluxes    &   mean & median &      $R^2 > 0.90$ & significant \\
\midrule
\multirow{3}{*}{Train} & 0.0 &  0.711 &  0.993 &  63.8 &  70.1 \\
           & 10.0 &  -6.89 &  0.56 &  12.2 &  19.7 \\
           & 50.0 &  -38.9 &  0.604 &  4.98 &  6.45 \\
\cline{1-6}
\multirow{3}{*}{Test} & 0.0 &  0.666 &  0.977 &  59.4 &  67.2 \\
           & 10.0 &  -4.7 &  0.33 &  6.67 &  15 \\
           & 50.0 &  -29.2 &  0.431 &  3.16 &  6.56 \\
\bottomrule
\end{tabular}
\end{table}

% \begin{figure}[!ht]   % Figure 2    (4?)
% \centering
% %\includegraphics[width=\columnwidth]{figures/perf_vs_1000rand.pdf}
%     \caption{Fit of fluxes predicted by the model 
%     using the recovered goal reaction, assessed by $R^2$
%     (coefficient of variation). Asterisks denote
%     significantly higher $R^2$ compared to 
%     1,000 fluxes each computed using a randomly
%     shuffled goal reaction ($P \leq 0.05$).
%     Columns correspond to experiments run with
%     different fractions of (randomly selected)
%     missing flux measurements.
%     \label{fig:perf_signif}
%     }
% \end{figure}

\begin{table}[htb]   % Table 3
    \centering
    \caption{Goal reaction recovery accuracy for 75 BiGG models.
    \label{tab:objrec}}
\begin{tabular}{rlrrrr}
\toprule
\% missing     &       Metric  & min &   max &    mean &  median \\ 
fluxes               \\ 
\midrule
\multirow{2}{*}{0.0} & Pearson &  0.021 &  1 &  0.96 &  1 \\
     & Spearman &  0.1 &  0.76 &  0.48 &  0.48 \\
\cline{1-6}
\multirow{2}{*}{10.0} & Pearson &  -0.74 &  1 &  0.063 &  -0.023 \\
     & Spearman &  -0.12 &  0.9 &  0.59 &  0.63 \\
\cline{1-6}
\multirow{2}{*}{50.0} & Pearson &  -0.92 &  0.82 &  -0.032 &  0.0012 \\
     & Spearman &  -0.27 &  0.76 &  0.38 &  0.4 \\
\bottomrule
\end{tabular}
\end{table}

%--------------------------------------------------
\subsection{Effect of multiple data types}
%--------------------------------------------------

%Real metabolic flux experiments measure
%a limited portion of the metabolic network---typically
%reactions in central metabolism \cite{antoniewicz2015methods} 
%(<$100$ reactions).
%
%We find that estimating a goal reaction
%from such limited measurements is difficult.
%Therefore, we investigate if including other data types 
%can improve reaction estimation.
%
Next, we investigate if including data types other than
metabolic fluxes can improve reaction estimation.
In particular, the concentration of over $95\%$ of cell proteins
and transcripts (RNA) can be measured \cite{schmidtNBT2016}.
These data types have been used to identify statistically
enriched biological functions \cite{hartTasks2015} but 
not to estimate reactions.
Thus, to include protein concentrations in our algorithm,
we extend \eqref{eq:FBA}, similar to \cite{FBAwMC07}. 
(Details in Supplement \ref{supp:protein}.)

We estimated goal reactions with three different data sets:
only metabolic fluxes, only proteins, and both.
To reflect realistic data, we include only
metabolic fluxes for central metabolism,
and include all proteins---this is achievable using
proteomics, or with RNA sequencing where RNA is used
to estimate protein concentrations.
We then use the estimated goal reaction to 
predict all metabolic fluxes and protein concentrations
(including values not seen by the algorithm)
in left-out conditions to assess the test performance.
%We conducted these experiments using
%small model of \textit{E. coli} metabolism \cite{bigg_models}.

When only metabolic fluxes are provided,
the prediction accuracy is low (mean $R^2= 0.281$) (Table \ref{tab:proteome}).
When only proteins are provided, 
the accuracy is even lower (mean $R^2=-16.4$).
One reason for the low accuracy is that protein concentrations
are not always directly proportional to flux---they only provide
an upper bound.
When both fluxes and proteins are provided,
the average test accuracy is $R^2=0.567$, which is
double the accuracy with only flux measurements.

%--------------------------------------------------
%%% Proteome
% 1 cond
\begin{table}[htb]   % Table 4
\centering
\caption{Test performance of flux and protein predictions
given different coverage of protein and flux measurements.
Coverage refers to the percentage of fluxes or proteins
that are measured and used as input to \ouralg{}.
\label{tab:proteome}}
\begin{tabular}{rr|rrr}
\toprule
\multicolumn{2}{c}{Coverage (\%)} & \multicolumn{3}{c}{$R^2$} \\
\cline{1-2} \cline{3-5}
  Flux & Protein     &    mean  & min  & max \\
\midrule
 57 &  100 &  0.567  &  0.108   &  0.77 \\
 57 &    0 &  0.281  &  0.225   &  0.322 \\
 0 &    100 & -16.375 &  -23.556 &  -11.293 \\
\bottomrule
\end{tabular}
\end{table}
%--------------------------------------------------

%--------------------------------------------------
% \begin{figure}[htb]   % Figure 3

% \caption{Accuracy given protein and flux measurements for large-scale models \label{tab:proteome_all}}
% \end{figure}
% %--------------------------------------------------

%--------------------------------------------------
\subsection{Multi-core speedup}
%--------------------------------------------------

\begin{figure}
    \centering
    \includegraphics[width=0.55\columnwidth]{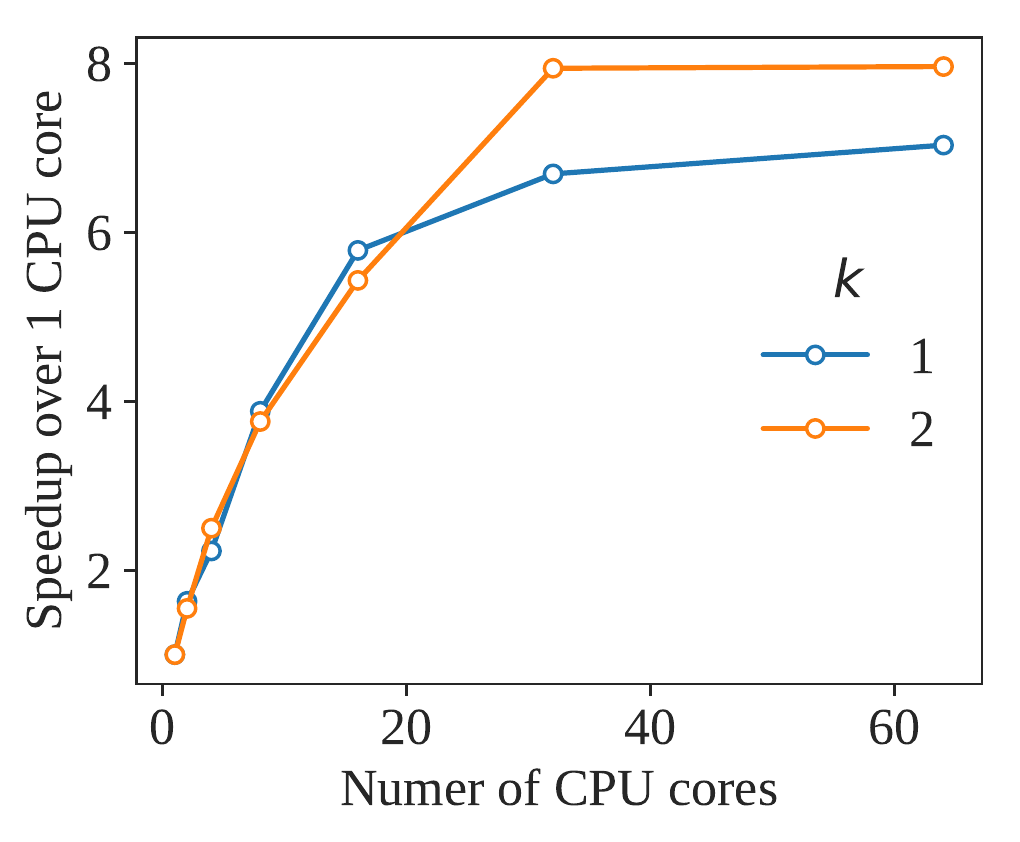}
    \caption{ 
        Speedup with multiple CPU cores for number of conditions
        $\numCond=1$ and $2$
        using the fine-grained implementation of \ouralg{}.
        \label{fig:cpuspeed}
    }
\end{figure}

Problem \eqref{eq:biLevelProg} can grow rapidly in size
and can benefit from parallel computing. BOSS cannot exploit parallel computational resources, but \ouralg{} can.
% since the numbers of variables and constraints
% are proportional to the number of conditions
% $\numCond$, except for the coupling variable
% $\newStoi$ that is shared across conditions.
%
% Furthermore, the number of nonconvex bilinear constraints
% is equal to the $\numCond$.
% With $\numCond=1$ condition, using a typical large-scale
% metabolic network like iML1515, we have 11,884 variables
% and 11,886 constraints. 
% With 100 conditions, 
% a realistic number for RNA sequencing data,
% we have over 1 million variables and constraints, of which 100 would be nonconvex.
%
We thus tested how \ouralg{} scales with multiple processors,
using the same metabolic network as in Section \ref{sec:compareboss}.
To maximize parallelism, we implemented a fine-grained
version of our algorithm, similar to \cite{hao2016fine}. The fact that we are using ADMM is important to achieve this.
%
%First, note that for small $\numCond$, we stack the
%$\numCond$ linear systems into one large linear system.
%Code profiling indicates that the majority of time is spent
%solving the linear system, so there is little room for 
%additional parallelism. 
%Alternatively, we can split $\numCond$ linear systems in parallel
%as independent proximal operators.
This implementation involves splitting node $\AQP$: 
one for every constraint.
%each row of
%$\blkLhsI \dummyI = \blkRhsI$, $\forall i \in \{1,\dots,\numCond\}$.
By combining this fine-grained version of $\AQP$
with $\ABD$ and $\ABI$, we have our fine-grained \ouralg{}.
This implementation of \ouralg{}
gave 1.6$\times$ speedup with 2 CPUs,
and 7$\times$ speedup with 64 CPUs (Fig.~\ref{fig:cpuspeed}).
With $\numCond=2$ the speedup was greater, 
achieving $8\times$ with 32 and 64 CPUs.

%We generated measured fluxes for $\numCond=64$ nutrient conditions,
%resulting in 
%642,325 variables and 642,453 constraints of which 64 are nonconvex.
% Update above for proteome
%We then solved \ouralg{} using $1$ to $64$ CPU cores.
%In order to improve scalability, we split node $\AQP$ into
%$\numCond$ proximal operators.

% NOTE BELOW
In these experiments, we did not perform preconditioning or adaptive penalty
to prevent factors besides CPU count from affecting speed.
To parallelize our code, we used OpenACC and the PGI compiler 18.10.
These tests were performed on Google Compute Engine
virtual machines with 32-core (64-thread) Intel Xeon 2.30 GHz processors 
and 64 GB RAM.

%--------------------------------------------------
%--------------------------------------------------

%==================================================
\section{Conclusion}
%==================================================

In this work, we addressed the problem of estimating
cellular goal reactions from measured fluxes in
the context of constraint-based modeling of cell metabolism.
This problem was previously addressed by the BOSS algorithm \cite{boss2008}, which successfully demonstrated the viability
of this approach on a model of \textit{Saccharomyces cerevisiae} central metabolism composed of 60 metabolites and 62 reactions.
Here, we developed a new method that extends BOSS and demonstrates
its performance on
%the latest genome-scale model of \textit{E. coli}
%metabolism (iML1515), consisting of 1877 metabolites and 2712 reactions \cite{iML1515}.
75 metabolic networks having up to 10,600 reactions and 5,835 metabolites. 
%The median model size was $2,727$ reactions and $1,942$ metabolites.

Our method successfully estimated goal reactions that
enabled accurate prediction of metabolic fluxes in new 
nutrient environments
(median coefficient of determination, $R^2=0.98$).
%which when maximized, led to flux predictions that 
%correlate with
%the provided training fluxes (coefficient of determination, $R^2$ of 0.99 to 1).
Furthermore, the stoichiometric coefficients of the estimated reactions
matched those of the ground truth reactions that were used to generate
the training fluxes (median Pearson correlation, $r = 0.96$).

As with the original BOSS, our algorithm involves a
nonconvex optimization problem with bilinear constraints.
%Using the iML1515 genome-scale model,
%the overall nonconvex optimization problem (for one condition) had 
%15,644 variables and
%22,112 constraints.
%The optimization problem for one condition ($\numCond=1$) had
%To solve this problem, we applied ADMM \cite{bento2013message,derbinsky2013improved}.
ADMM is an efficient approach for solving problems with many bilinear constraints, 
and more generally, nonconvex quadratic constraints \cite{huang2016consensus}.
For our problem, the number of bilinear constraints can be large, 
as it increases with the number of conditions in the data.
%[LY] 05/16/2019:
Besides scalability, \ouralg{} allows for regularization,
modularity---enabling extensions, and 
can use multiple data types (fluxes and protein concentrations),
which improves prediction accuracy.
%

%Thus, ADMM is the approach we use here.
%which found feasible solutions within minutes for the genome-scale model.
% Our ADMM implementation involves separation of the 
% objective function and constraints into a set of 
% proximal operators for which closed form solutions were available.
% The bilinear constraints, in particular, had a closed form
% solution involving solving a quartic equation.

Genome-scale metabolic network models have been invaluable
in numerous studies including infectious disease and
cancer metabolism \cite{bordbarCOBRA14}.
The success of these studies depends on the 
availability of a reaction representing cellular goals.
For many important cellular systemss, such as human tissue
or microbiomes, such reactions are still challenging
to determine experimentally \cite{feist2016cells}.
Our study shows that a data-driven approach for 
cellular goal estimation is promising.
This approach is most effective when we have 
high coverage flux measurements or
complementary data types, such as flux measurements
for parts of the network and protein or RNA abundance
measurements for most of the metabolic genes.
\ouralg{} is thus suited for 
analyzing large gene expression data sets,
e.g., from cancer tissue.
\ouralg{} is available at \url{https://github.com/laurenceyang33/cellgoal}.

%For certain organisms under specific environments,
%well-tested objective functions are available: 
%maximizing biomass synthesis or ATP yield, or minimizing 
%reactive oxygen species production.
%However, these objectives do not apply under many scenarios of 
%practical relevance \cite{feist2016cells},
%including specialized functions of mammalian tissues
%or organelles, and microbes under stress states or in a community.
%For these scenarios, data-driven estimation
%of \textit{de novo} cellular objectives offers
%a systematic approach for modeling cell metabolism.

\section{Acknowledgements}
This work was supported by the NIH grants 2R01GM057089
and 1U01AI124302, NSF grant IIS-1741129,
and Novo Nordisk Foundation grant NNF10CC1016517.
%and a GPU grant by NVIDIA Corporation.

% The next two lines define the bibliography style to be used, and the bibliography file.
\bibliographystyle{ACM-Reference-Format}
\bibliography{references}

% 
% If your work has an appendix, this is the place to put it.
\appendix

\section{Details for reproducibility}

\subsection{Model formulation including proteins \label{supp:protein}}
Our model formulation for including protein concentrations is
based on \cite{FBAwMC07} and \cite{lloyd2018cobrame}.
First, we denote $\prot$ as the vector of protein concentrations,
whose length is the number of proteins in the metabolic network.
Each protein has a molecular weight, which is stored in $\pweight$.
One or more proteins combine to form enzyme complexes, $\enzyme$,
which are ultimately the molecules that catalyze metabolic reactions.
The rate of a metabolic reaction is limited by the 
catalytic efficiency of an associated enzyme and that enzyme's
concentration. Since multiple enzymes can sometimes catalyze
the same reaction, we have a matrix of catalytic efficiencies,
$\keff$ mapping reactions to all possible enzymes.
The total protein mass of a cell, $\ptot$, can be measured, 
and this quantity imposes an upper bound on the sum of 
individual protein concentrations multiplied by 
their molecular weights, $\pweight$.
Finally, each protein can be part of one or more enzyme
complexes, and each enzyme complex is comprised of a fixed number
of a particular protein. This protein to enzyme mapping is
encoded in the matrix $\cplxstoi$.

By adding these relationships to $\eqref{eq:FBA}$, we
have the following linear problem:
\begin{equation}\label{eq:pcfba}
\begin{split}
\max_{\flux,\prot,\enzyme}    \quad& \objc\T \flux + \objp\T \prot \\
\text{\ subject to\ } \quad& \stoi \flux = \bal, \\
                      & \flux - \keff \enzyme + \slack^{(1)} = 0,      \\
                      & \pweight\T \prot + \slack^{(2)} = t,  \\
                      & \cplxstoi \enzyme - \prot + \slack^{(3)} = 0, \\
                      & \lowB \leq \flux \leq \uppB, \\
                      & \prot, \enzyme \geq 0, \\
                      & \slack^{(j)} \geq 0, \forall j\in \{1,2,3\},
                      %& \slack^{(1)},\slack^{(2)},\slack^{(3)}\geq 0,
\end{split}
\end{equation}
where
$\objp$ is the cell's relative importance of different proteins,
and $\slack^{(j)}$ are non-negative slack variables.

By defining $\dummy=\{ \flux, \prot, \enzyme, \slack^{(1)},\slack^{(2)},\slack^{(3)}\}$, 
$$
\bar{\stoi} = 
\begin{bmatrix}
    \stoi & 0 & 0 & 0 & 0 & 0 \\
    I & 0 & -K & I & 0 & 0 \\
    0 & -I & E & 0 & 0 & I 
\end{bmatrix},
$$
$\objpc = \{\objc, \objp\}$,
$\pclowB = \{\lowB, 0, 0, 0, 0, 0 \}$,
$\pcuppB = \{\uppB, +\infty,+\infty,+\infty\,+\infty,+\infty \}$,
and $\pcbal = \{\bal,0,t,0 \}$,
we can write \eqref{eq:pcfba} as
\begin{equation}
\max_{\dummy} \objpc\T \dummy \text{\ subject to\ } \pcstoi \dummy = \pcbal ,
\  \pclowB \leq \dummy \leq \pcuppB,
\end{equation}
which is \eqref{eq:FBA} with a change of variables.
Therefore, \ouralg{} can be applied directly to 
the problem with proteins and fluxes.

The $\keff$ matrix is typically not known fully.
Here, we used a default value of $65$ second$^{-1}$, 
which are in units of enzyme turnover rate and
represents an average catalytic efficiency for
all enzymes in \textit{E. coli}.
Values for matrix $\cplxstoi$ were determined
from the gene-protein-reaction mappings that are
included with each metabolic model in the BiGG database
\cite{bigg_models}. Without additional information we assumed
that each complex requires one unit of each protein
in the gene-protein mapping. 
For $\ptot$, we chose a default
value of 0.30, indicating that the model
accounts for $30$\% of the cell's protein mass.
For molecular weights $\pweight$, we used a default
value of 77 kDa for every protein, based on the 
average molecular weight of proteins in \textit{E. coli}.
(Note that $\pweight$ can also be determined directly 
from the protein's amino acid sequence.)

%Then, we can provide both flux and protein measurements,
%as $\{\fluxMes, \protMes \}$.

%, such that the \ouralg{} formulation becomes
%
% \begin{align}
% \min_{\{\dummyI\}^\numCond_{i=1},\newStoi} \ & \frac{1}{\numCond}
%     \sum_{i=1}^\numCond \norm{ \indFI \dummyI - \dummyMesI }_2^2 + 
%     \sparse \norm{ \newStoi }_1 \label{eq:protbiLevelProg}\\
% \text{ subject to} \ & \dummyI \in \mathcal{C}_i(\newStoi), \nonumber
% \end{align}
%
%where $\dummyMesI = \{\fluxMesI, \protMesI \}$

%has the same formulation for 
%the problem involving proteins and fluxes as it does for
%fluxes alone.

%\subsection{Software and hardware} 
\subsection{Online Resources}
Code and documentation for \ouralg{} are available at
\url{https://github.com/laurenceyang33/cellgoal}.
Detailed installation instructions are available there.
Briefly, users will need a Fortran compiler, CMake,
and a linear system solver. We tested our algorithm with both
UMFPACK, which is part of SuiteSparse, and
MA57 \cite{duff2004ma57} from HSL \cite{hsl2013}.
For parallel computing features, users will need
OpenACC and the PGI Compiler. We have tested all code
using PGI Community Edition version 18.10.
In the code repository, we include test suites to run \ouralg{} 
on metabolic network models that users can download
from the BiGG database \cite{bigg_models}.

% [LY] 05/17/2019: Added details on time
\subsection{Details on solution time} \label{apps:other_acc}
\begin{figure}[!hbt]
\centering
\includegraphics[width=\columnwidth]{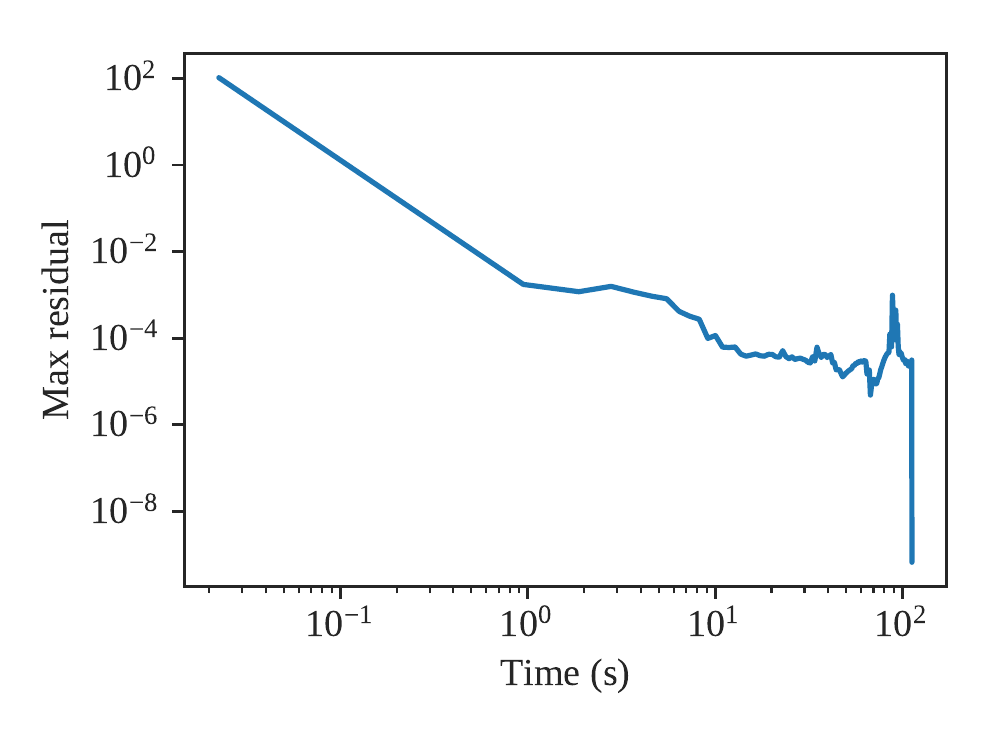}
\caption{
%Detailed run times for \ouralg{}
%when solving the problem in Fig.~\ref{fig:compareboss}b and d.
Maximum of primal and dual residuals versus
the wall time of running \ouralg{}. 
\label{fig:time}
}
\end{figure}

We report details on the run time for \ouralg{}
when solving the problem shown in Fig.~\ref{fig:compareboss}.
Solutions having modest accuracy are found quickly, as is
expected for ADMM (Fig.~\ref{fig:time}).
%

% Nam id fermentum dui. Suspendisse sagittis tortor a nulla mollis, in pulvinar ex pretium. Sed interdum orci quis metus euismod, et sagittis enim maximus. Vestibulum gravida massa ut felis suscipit congue. Quisque mattis elit a risus ultrices commodo venenatis eget dui. Etiam sagittis eleifend elementum. 

\end{document}